\pdfoutput=1
\documentclass[11pt]{article}

\usepackage{jheppub,booktabs}
\usepackage{amsmath,amssymb,graphicx} 
\usepackage{epsf}
\usepackage{epsfig} 

\newcommand{\ssw}{s^{2}_W}
\newcommand{\cw}{c_W}

\newcommand{\re}{{\rm Re}}
\newcommand{\im}{{\rm Im}}

\title{The LHC potential of Vector-like quark doublets} 

\author[a]{Giacomo Cacciapaglia,} 
\author[a,1]{Aldo Deandrea,
   \note{also Institut Universitaire de France, 103 boulevard
     Saint-Michel, 75005 Paris, France}} 
\author[b]{Naveen Gaur,}
\author[c]{Daisuke Harada,}
\author[d,e]{Yasuhiro Okada,}
\author[f]{Luca Panizzi}

\affiliation[a]{Universit\'e de Lyon, France; Universit\'e Lyon 1,
  CNRS/IN2P3, UMR5822 IPNL,\\ F-69622 Villeurbanne Cedex, France.} 
\affiliation[b]{Department of Physics, Dyal Singh College (University
  of Delhi), Lodi Road, New Delhi - 110003, India}   
\affiliation[c]{Zhejiang Institute of Modern Physics and Department of
  Physics, Zhejiang University, Hangzhou, Zhejiang 310027, China} 
\affiliation[d]{KEK Theory Center, Institute of Particle and Nuclear
  Studies, KEK, 1-1 Oho, Tsukuba, Ibaraki 305-0801, Japan} 
\affiliation[e]{Department of Particle and Nuclear Physics, Graduate
  University for Advanced Studies (Sokendai), 1-1 Oho,  
Tsukuba, Ibaraki 305-0801, Japan}
\affiliation[f]{Dipartimento di Fisica, Universit\`a di Pisa and 
  INFN, Sezione di Pisa, Largo Pontecorvo 3, I-56127 Pisa, Italy} 

\emailAdd{g.cacciapaglia@ipnl.in2p3.fr}
\emailAdd{deandrea@ipnl.in2p3.fr}
\emailAdd{gaur.nav@gmail.com}
\emailAdd{dharada@zju.edu.cn}
\emailAdd{yasuhiro.okada@kek.jp}
\emailAdd{lcpnzz@gmail.com}

\abstract{The existence of new vector-like quarks is often predicted
  by models of new physics beyond the Standard Model, and the
  development of discovery strategies at colliders is the object of an
  intense effort from the high-energy community. Our analysis aims at
  identifying the constraints on and peculiar signatures of simplified
  scenarios containing \textit{two} vector-like quark doublets mixing
  with \textit{any} of the SM quark generations. This scenario is a
  necessary ingredient of a broad class of theoretically motivated
  constructions. We focus on the two charge $2/3$ states
  $t_{1,2}^\prime$ that, due to their peculiar mixing patterns,
  feature new production and decay modes that are not searched for at
  the LHC: single production of the heavier state can dominate over
  the light one, while pair production via electroweak interactions
  overcomes the QCD one for masses at the TeV scale. }

\begin{document}

\begin{flushright}
  KEK-TH-2042    \\
  LYCEN-2018-04
\end{flushright}
\maketitle
\flushbottom


\section{Introduction}                          
\label{sec:introduction} 

The search for particles beyond the Standard Model (SM) is one of the
main goals of the Large Hadron Collider (LHC). In the midst of the Run
II, a new range of energies is being explored, thus playing a crucial
role in finding new phenomena or setting bounds on various aspects of
New Physics (NP) models. The progress in the understanding of the
Higgs sector via the Higgs coupling measurements at the LHC is also a
major advance in the exploration of NP, as it allows to test the
extensions of the SM either in new channels at colliders or to
envisage new complementary ways to explore the presently explored
final states. Among the many NP states searched for at the LHC,
vector-like quarks (VLQs) play a prominent role in terms of
experimental effort. A large number of searches have been performed by
both ATLAS and CMS, exploring pair and single production of VLQs in a
wide range of possible final states and signatures. No evidence of
their existence has been observed so far, giving rise to mass bounds
in the TeV range. The precise values depend on assumptions on the
allowed decay channels and particular mixing with the SM quarks, and
the bounds are overall robust if the mixing of the VLQs is mainly to
3rd generation quarks.  

The fact that VLQs are the object of such an extensive exploration did
not happen by chance: in fact, they are predicted or suggested by a
large number of extensions of the SM, especially in relation with the
top quark. As examples, VLQs appear as top partners in composite Higgs
models~\cite{Agashe:2004rs,Contino:2006qr,Giudice:2007fh,Matsedonskyi:2012ym},
extra-dimensional
models~\cite{Antoniadis:1990ew,Antoniadis:2001cv,Csaki:2003sh,Hosotani:2004wv,Cacciapaglia:2009pa},
gauge-Higgs models \cite{Hosotani:1983xw}, models with gauge coupling
unification \cite{Choudhury:2001hs,Panico:2008bx}, little Higgs models
\cite{ArkaniHamed:2002qx,ArkaniHamed:2002qy,Schmaltz:2005ky} and
models with an extended custodial symmetry
\cite{Agashe:2006at,Chivukula:2011jh}. Typically, the experimental
searches have been based on simplifying assumptions guided by the
expectations in specific models, like mixing with the third generation
of SM quarks and decays into a $W$, $Z$ or Higgs plus a top or bottom
quark
\cite{delAguila:2000rc,AguilarSaavedra:2005pv,Anastasiou:2009rv,AguilarSaavedra:2009es,Cacciapaglia:2010vn,Marzocca:2012zn,DeSimone:2012fs,Falkowski:2013jya,Aguilar-Saavedra:2013wba,Ellis:2014dza}.
In general, however, the mixing with the first and second SM
generations needs to be considered
\cite{Atre:2011ae,Cacciapaglia:2011fx,Okada:2012gy,Buchkremer:2013bha,Delaunay:2013pwa,Barducci:2014ila},
and a few LHC searches are also available
\cite{Aad:2011yn,Sirunyan:2017lzl}. Furthermore, decays into a non-SM
boson
\cite{Serra:2015xfa,Brooijmans:2016vro,Aguilar-Saavedra:2017giu,Chala:2017xgc,Bizot:2018tds}
or Dark Matter
\cite{Anandakrishnan:2015yfa,Kraml:2016eti,Moretti:2017qby,Balkin:2017aep,Chala:2018qdf}
are recently receiving increasing attention. 

Apart from the specific set-up required by these models, it is
interesting to study VLQs in a more general context, and we consider
this possibility in the following. A common situation in NP models is
the presence of extended global symmetries that require several VLQ
multiplets, which remain close in mass. These multiplets mix with the
SM quarks and among each other via Yukawa-type interactions of the
Higgs field. This in turn affects the tree-level and loop-level bounds
on masses and coupling strengths, modifying the results and the
expectations obtained in simplified analyses. In the present work we
further generalise the analysis we performed in
\cite{Cacciapaglia:2015ixa} by considering general structures and
mixing of more than one VLQ multiplet mixing s with the three SM quark
generations. We take into account updated bounds both from direct
searches, Higgs physics and Electroweak Precision Tests (EWPT). In
particular we shall focus on the case of non-degenerate SU(2)$_L$
doublets, which is of particular interest for model building with
extended custodial symmetry. Furthermore, these multiplets feature a
cancellation at low energy that relaxes the typically very strong
bounds coming from precision electroweak observable.  

Our main objective is to explore signatures that are characteristic of
this specific bi-doublet configuration, and that can be used to
distinguish models containing these multiplets from other generic VLQ
models. We will identify configurations where the observation of the
heavier VLQs is favoured with respect to the lightest one of the
multiplets, and specific decay patterns for the charge $2/3$
VLQs. Finally, we point out the importance of pair production of two
VLQs via electroweak interactions, which can dominate over the QCD
pair production for large (allowed) mixing. This feature was, to the
best of our knowledge, first noted in
Ref.~\cite{Cacciapaglia:2009cu}. 

The paper is organised as follows: in Section 2 we recall the
structures and properties of VLQ doublets, their relevance in well
known models and the typical cases in which they feature cancellations
that allow to reduce their impact on low-energy observable. In Section
3 we discuss indirect bounds from EWPT, tree level and loop level
contributions to the $Z$ and Higgs couplings, and bounds from current
direct searches. In Section 4 we discuss the main new features that
lead to novel signatures at the LHC, before presenting our conclusions
in Section 5. 


\section{Vector-like multiplets: models with two doublets}                  
\label{sec:model}
The general description of the first few VLQ multiplets is given in
\cite{Cacciapaglia:2015ixa}, where they are classified in terms of
both their quantum numbers and their particle content (multiplets
containing  top partners,  bottom partners, or both). In addition to
partners of the standard quarks, these multiplets may contain other
exotic charged VLQ particles. The VLQ multiplets that can mix with SM
quarks and a SM (or SM-like) Higgs boson have been studied, rather
extensively, in the literature
\cite{delAguila:1982fs,delAguila:2000rc,Aguilar-Saavedra:2013wba,Cacciapaglia:2010vn,Cacciapaglia:2011fx,Okada:2012gy,Buchkremer:2013bha}.
In the following we focus on the specific case of VLQ doublets, as it
is of particular importance in various extensions of the SM with an
extended custodial symmetry (see, {\it e.g.}, Refs
\cite{Agashe:2006at,Chivukula:2011jh}). The doublets we consider in
the following are $\left(\begin{array}{c}U_1 \\ D_1
  \end{array}\right)_{1/6}$ and $\left( \begin{array}{c} X_2^{5/3} \\
    U_2 \end{array} \right)_{7/6}$\,, where the subscript number
represents the hypercharge of the multiplet, and the exotic state
$X^{5/3}$ has electromagnetic charge $+5/3\ e$. The presence of VLQ
multiplets generically allows to add new Yukawa interactions between
the VLQ multiplets and the SM quarks, or among VLQ multiplets,
mediated by scalar fields from the Higgs sector. Gauge invariance
requires that new VLQ doublets couple with the SM right-handed
singlets (if the Higgs sector is not modified). For VLQ multiplets
with the same quantum numbers as the SM quarks, a direct mass mixing
can be written down but it is not physical, as it can be removed
redefining the fields corresponding to the SM and VLQs. A description
of the Lagrangian terms and mass matrices for scenarios with two
doublets can be found in Appendix~\ref{app:multiplets}, where we also
include the case of a doublet with hypercharge $-5/6$. The latter
features an exotic charged bottom-partner, and we will consider its
phenomenology in a follow-up work.

A detailed account of the Yukawa structure and mixing patterns can be
found in \cite{Cacciapaglia:2015ixa}. In the remaining of this section
we will consider, in detail, the relation between the general
formalism we use in this paper and composite (pseudo-Nambu-Goldstone)
Higgs models. 

\subsection{Relation to composite top partners}
\label{sec:composite}

In models of composite top partners, where the elementary tops pick up
a mass via mixing with composite operators~\cite{Kaplan:1991dc},
bi-doublets like the ones we consider in this paper arise
naturally. This is due to the fact that the symmetries of the
composite sector need to include the full custodial
SO(4)$\sim$SU(2)$_L \times$SU(2)$_R$ symmetry of the Higgs
sector~\cite{Georgi:1984af}, and top partners embedded in a bi-doublet
are preferred by the absence of dangerous tree level corrections to
the $Z$ couplings to the left-handed bottom
quarks~\cite{Agashe:2006at}. The main difference between the composite
case and the Lagrangian we adopted in Eq.~(\ref{eq:LV-SM}) (for the
case relevant for the top mass generation) is twofold: on the one
hand, in the effective Lagrangian for partially composite
tops~\cite{Marzocca:2012zn}, the elementary fields corresponding to
the SM tops do not couple directly to the Higgs boson but mix linearly
with the composite operators via a mass term generated by the
condensate; on the other hand, the Higgs field enters non-linearly in
the couplings, thus higher order couplings are implicitly included. 

To establish a bridge between our study and models with partially
composite tops, we detail here the correspondence between our
parameters and the ones of a model based on the symmetry breaking
SO(5)/SO(4) (so-called minimal composite Higgs), where the top
partners are allowed to transform as a ${\bf 4}$ of the unbroken
symmetry SO(4)~\cite{Agashe:2006at,Contino:2006qr}. This discussion is
actually valid for any symmetry breaking pattern, as long as an
unbroken custodial SO(4) is contained in the unbroken subgroup. We
will follow the notation of Ref.~\cite{Cacciapaglia:2015dsa}, where
the mass mixing in the effective Lagrangian description reads: 
\begin{multline}
\mathcal{L}_{CHM} \supset - M_4\ (\bar{T}_L T_R + \bar{B}_L B_R +
\bar{X}_{5/3 L} X_{5/3 R} + \bar{X}_{2/3 L} X_{2/3 R}) + \\ 
- y_{L4} f\ \left(\bar{b}_L B_R + \cos^2 \frac{\theta}{2} \ \bar{t}_L
  T_R + \sin^2 \frac{\theta}{2}\ \bar{t}_L X_{2/3 R} \right) -
\frac{y_{R4} f \sin \theta}{\sqrt{2}} (\bar{T}_L t_R - \bar{X}_{2/3 L}
t_R) + \mbox{h.c.} 
\end{multline}
where $(T,B)$ and $(X_{5/3}, X_{2/3})$ are the two doublets that share
a common mass $M_4$; $f$ is the decay constant of the pions in the
composite sector (including the Higgs boson) and the angle $\theta$
parameterises in a non-linear way the Higgs vacuum expectation value
(VEV), such that $v = f \sin \theta$. Note that the SM elementary
doublet $(t,b)$ mixes with the composite doublet with strength $y_{L4}
f$ not suppressed by the Higgs VEV, so that we can remove this term by
redefining: 
\begin{equation}
t_L = s_{\theta L} U_{1L} + c_{\theta L} u_L^3\,, \quad T_L =
c_{\theta L} U_{1L} - s_{\theta L} u_l^3\,, \qquad s_{\theta L} = \sin
\theta_L = \frac{y_{L4} f}{\sqrt{M_4^2 + y_{L4}^2 f^2}}\,, 
\end{equation}
and analogously $b_L = s_{\theta L} D_{1L} + c_{\theta L} d_L^3$ and
$B_L = c_{\theta L} D_{1L} - s_{\theta L} d_l^3$. Upon identifying the
fields $t_R \equiv u_R^3$, $T_R \equiv U_{1R}$, $X_{2/3} \equiv U_2$
and $X_{5/3} \equiv X^{5/3}$, at leading order in the Higgs VEV the
parameters in our Lagrangian~(\ref{eq:LV-SM}) match the composite ones
as follows: 
\begin{equation}
M_1 = \sqrt{M_4^2 + y_{L4}^2 f^2}\,, \quad M_2 = M_4\; (< M_1)\,, 
\end{equation}
and
\begin{equation}
\tilde{m}_{33}^{\rm up} = - \frac{y_{R4} f \sin \theta}{\sqrt{2}}
s_{\theta L}\,, \quad y_{1u}^3 = \frac{y_{R4} f \sin \theta}{\sqrt{2}}
c_{\theta L}\,, \quad y_2^3 = \frac{y_{R4} f \sin \theta}{\sqrt{2}}\;
(> y_{1u}^3)\,. 
\end{equation}
The above formulas show that composite models indeed prefer masses for
the two doublets that are not equal (and in particular, the hierarchy
$M_2 < M_1$ is an outcome) as well as unequal Yukawa $y_2^3 >
y_{1u}^3$. 

Another interesting possibility, which has deserved attention in the
literature, is that the right-handed top component is itself a fully
massless composite state~\cite{Panico:2012uw,DeSimone:2012fs}. In this
case, a direct coupling of the left-handed elementary tops is allowed:
\begin{multline}
\mathcal{L}_{CtR} \supset - M_4\ (\bar{T}_L T_R + \bar{B}_L B_R +
\bar{X}_{5/3 L} X_{5/3 R} + \bar{X}_{2/3 L} X_{2/3 R}) + \\ 
- y_{L4} f\ \left(\bar{b}_L B_R + \cos^2 \frac{\theta}{2} \ \bar{t}_L
  T_R + \sin^2 \frac{\theta}{2}\ \bar{t}_L X_{2/3 R} \right) -
\frac{y_{Rt} f \sin \theta}{\sqrt{2}}\  \bar{t}_R t_L + \mbox{h.c.} 
\end{multline}
where we see that the coupling between the right-handed top and the
heavy doublets is replaced by a direct Yukawa with the light
left-handed top. The same rotation among doublets can be done as
before, now leading to the following identification of Yukawa
couplings: 
\begin{equation}
\tilde{m}_{33}^{\rm up} = \frac{y_{Rt} f \sin \theta}{\sqrt{2}}
c_{\theta L}\,, \quad y_{1u}^3 = \frac{y_{Rt} f \sin \theta}{\sqrt{2}}
s_{\theta L}\,, \quad y_2^3 = 0\,; 
\end{equation}
while the masses of the heavy doublets are the same as above.

\section{Constraints on the parameter space} 
\label{sec:ewbounds}

We examine, in the following, the scenario with two doublets of
hypercharge $1/6$ and $7/6$ respectively, each containing a charge
$2/3$ top partner, labeled as $U_{1,2}$ in the gauge eigenstate basis
and $t^\prime_{1,2}$ in the mass one, where
$m_{t_1^\prime}<m_{t_2^\prime}$. The relation between the masses of
$t_{1,2}^\prime$ and the Lagrangian parameters $M_{1,2}$ after the
diagonalisation of the mass matrix is described in Appendix
\ref{app:masses}. In the numerical study, we considered benchmark
values for the mass parameters in the Lagrangian ({\it i.e.} the VLQ
mass terms in the gauge eigenstates, before mixing) as follows:
$M_1=1000$ GeV and $M_2=\{1100,1200,1400\}$ GeV. We will thus show
selected results from those benchmarks. Note that in composite Higgs
models, one typically expects the opposite mass ordering for the
multiplets, however experimental bounds go rather in the opposite
direction as the bounds on the $X^{5/3}$ exotic charge member
(belonging to the second multiplet) are strong. We take, therefore,
benchmark points that take this fact into account and that allow  to
explore in a first step an overall lower range of masses which are
within immediate or close reach for the LHC. 

Indirect constraints on the spectrum and couplings of the VLQs arise
both at tree level, via modifications to the couplings of the $Z$ and
Higgs (and $W$) to the SM quarks \cite{delAguila:2000rc}, and at loop
level via contribution to the observable in the EWPT
\cite{Cacciapaglia:2010vn,Chen:2017hak} and loop-induced couplings of
the Higgs \cite{Bizot:2015zaa}. These constraints give a first
indication of the available parameter space that is still interesting
to further explore in direct searches at the LHC. Note, however, that
we are working under the assumption that the only light NP states are
the new VLQs. Thus, the effect of other states to EWPT is not taken
into account, and they may affect the results even if the new
particles are heavier than the VLQs. The reader should be wary,
therefore, that the loop-level indirect bounds should not be
considered as absolute bounds, but rather they should be taken as an
indication in models that contain other particles contributing to
these corrections. Tree level bounds, on the other hand, are more
solid as they arise directly from the mixing. 
%
\begin{figure}[h] 
\begin{center}
\epsfig{file=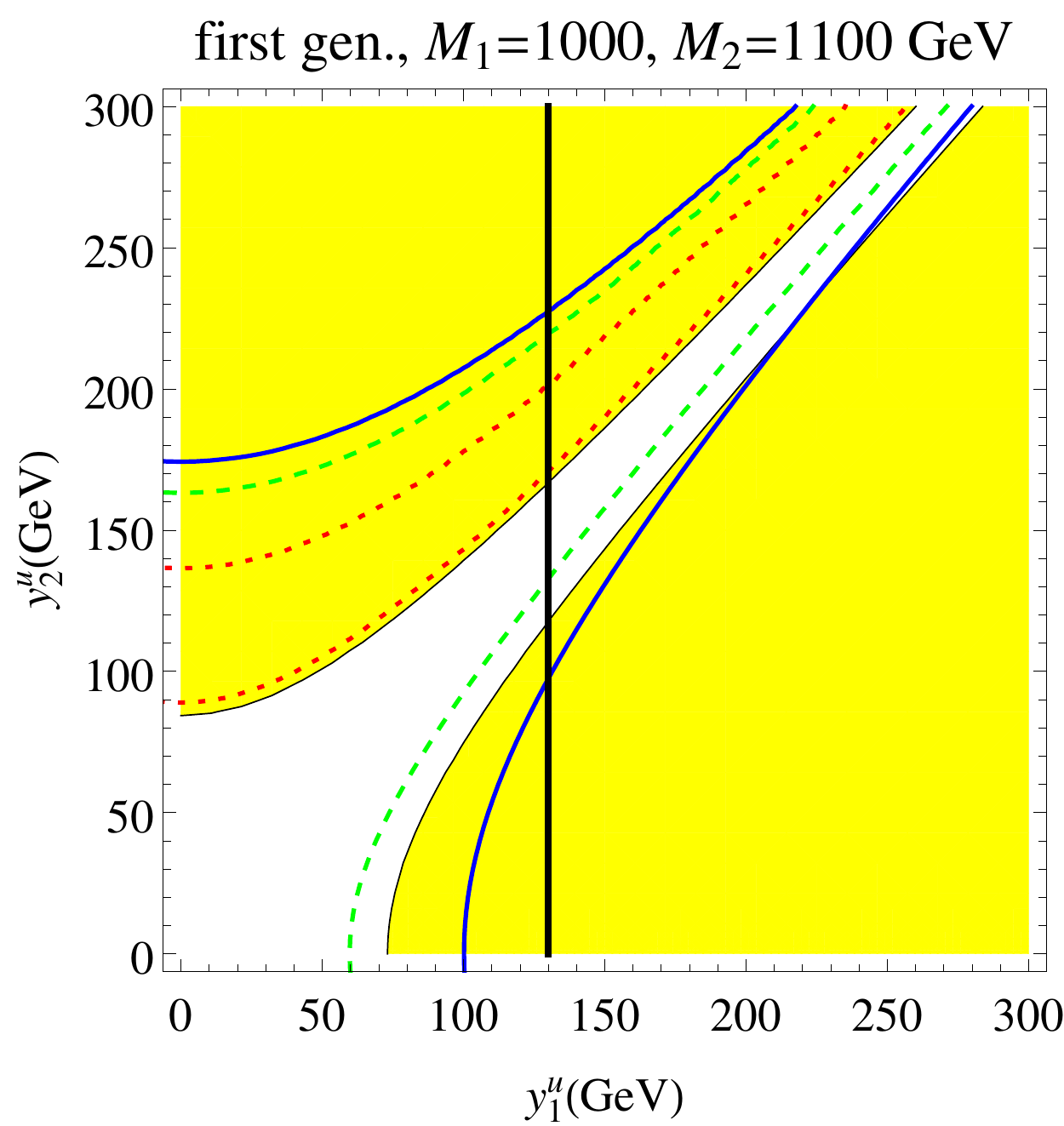,width=0.46\textwidth}\hfill
\epsfig{file=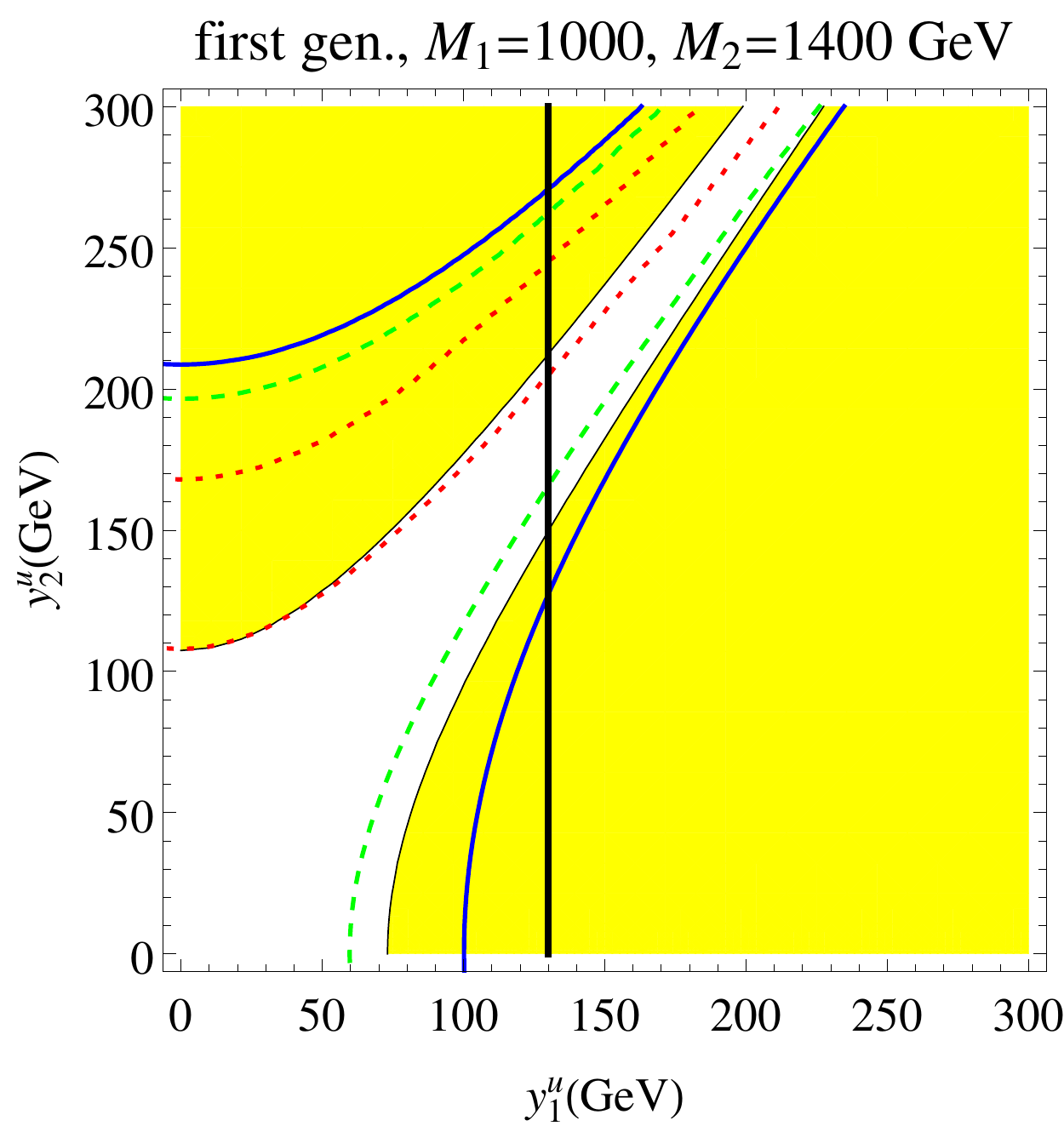,width=0.46\textwidth}\\[8pt]
\epsfig{file=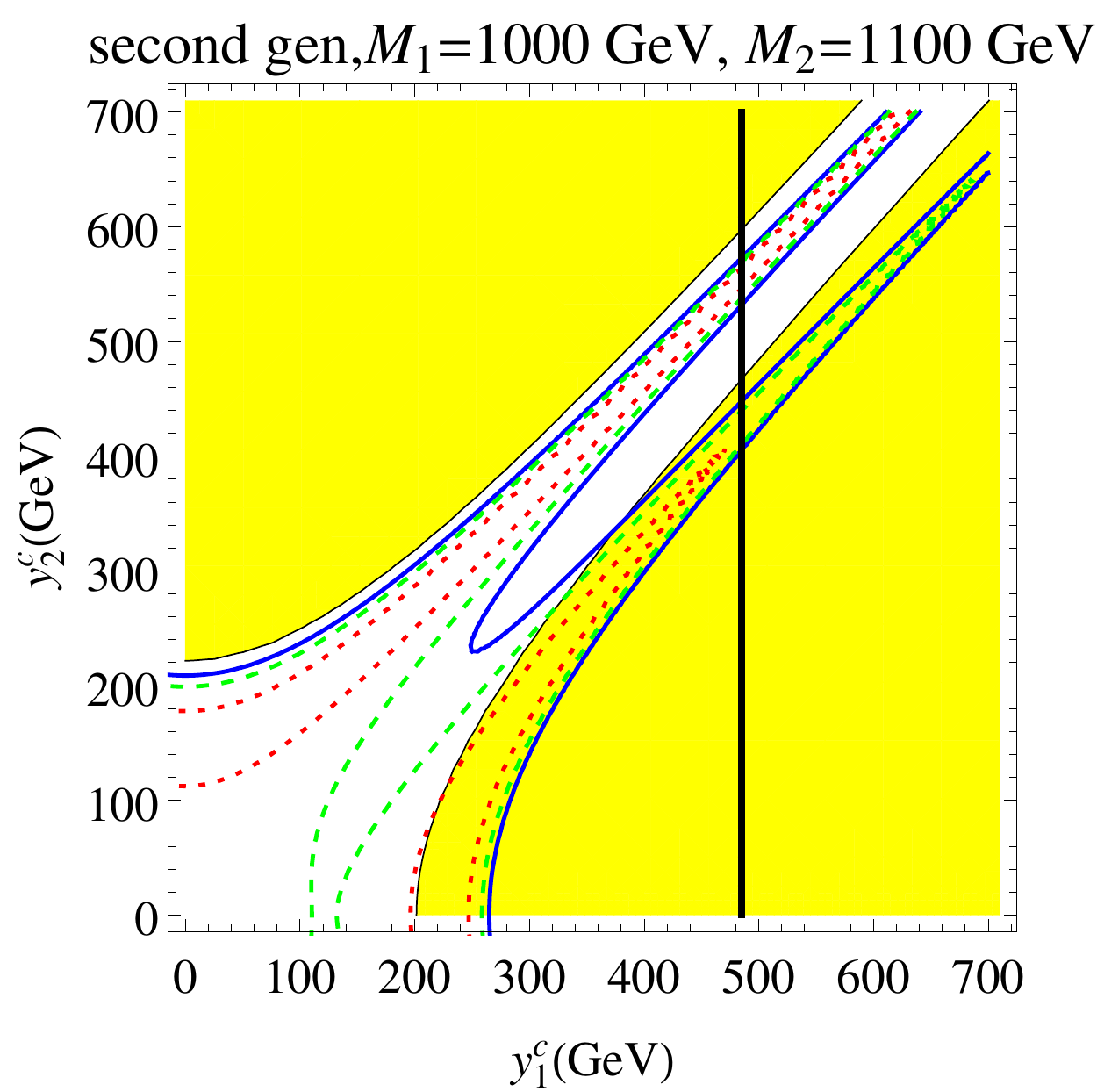,width=0.49\textwidth}\hfill
\epsfig{file=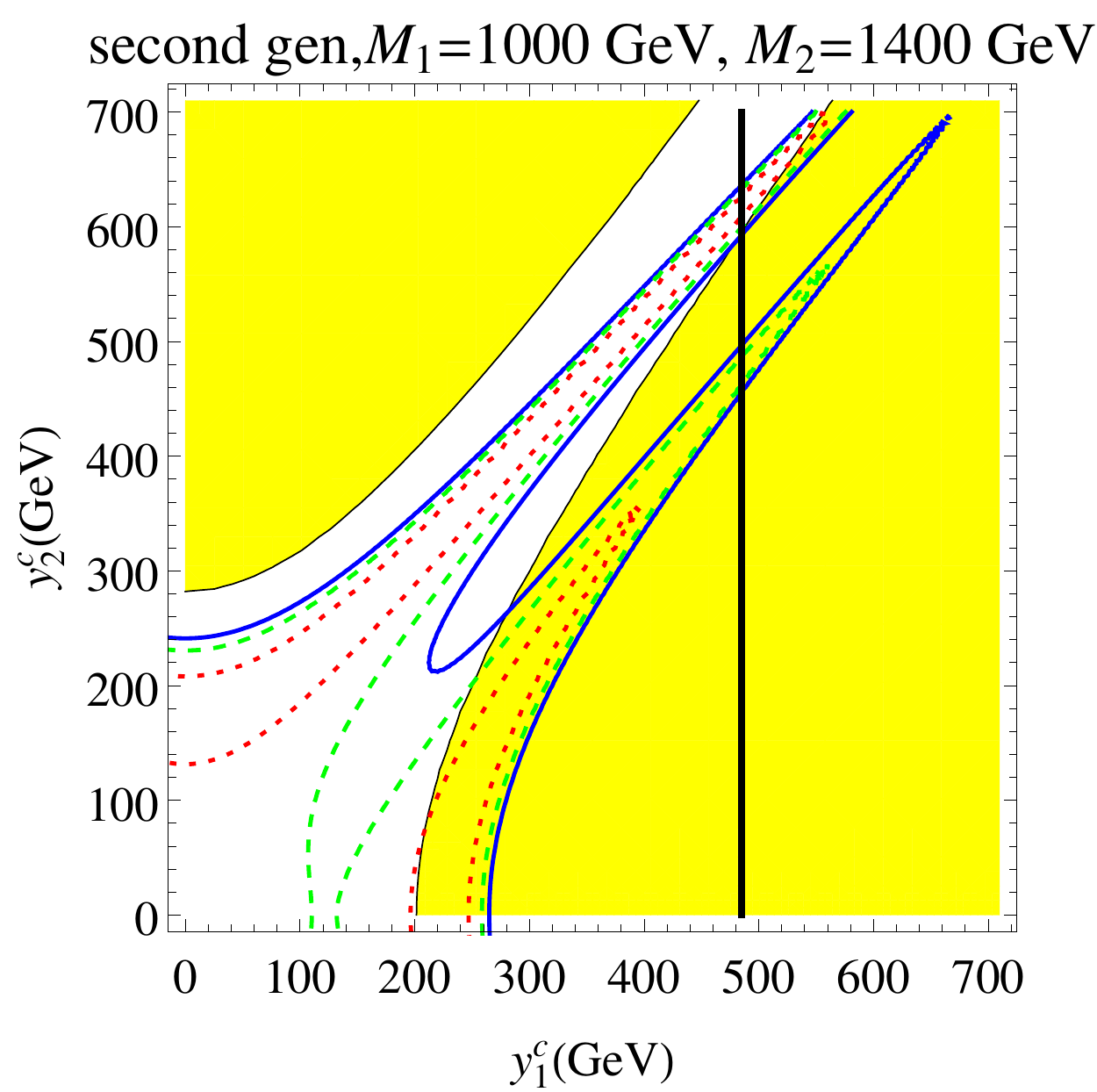,width=0.49\textwidth} 
\caption{\label{fig:ewptlight} Tree level (yellow area is excluded at
  3$\sigma$), EWPT (blue continuous line corresponds to the 3$\sigma$
  bound, green dashed to 2$\sigma$, red dotted to 1$\sigma$, the strip
  between the lines is allowed) and LHC single VLQ production bounds
  (vertical black line, excluded region on the right) in the case of
  mixing of two VLQ multiplets with the first (top panels) or second
  (bottom panels) SM quark generation. Plots on the left column
  correspond to benchmark masses $M_1=1000$ GeV and $M_2=1100$ GeV,
  while on the right to $M_1=1000$ GeV and $M_2=1400$ GeV.}    
\end{center}
\end{figure} 
%
\begin{figure}[h] 
\begin{center}
\epsfig{file=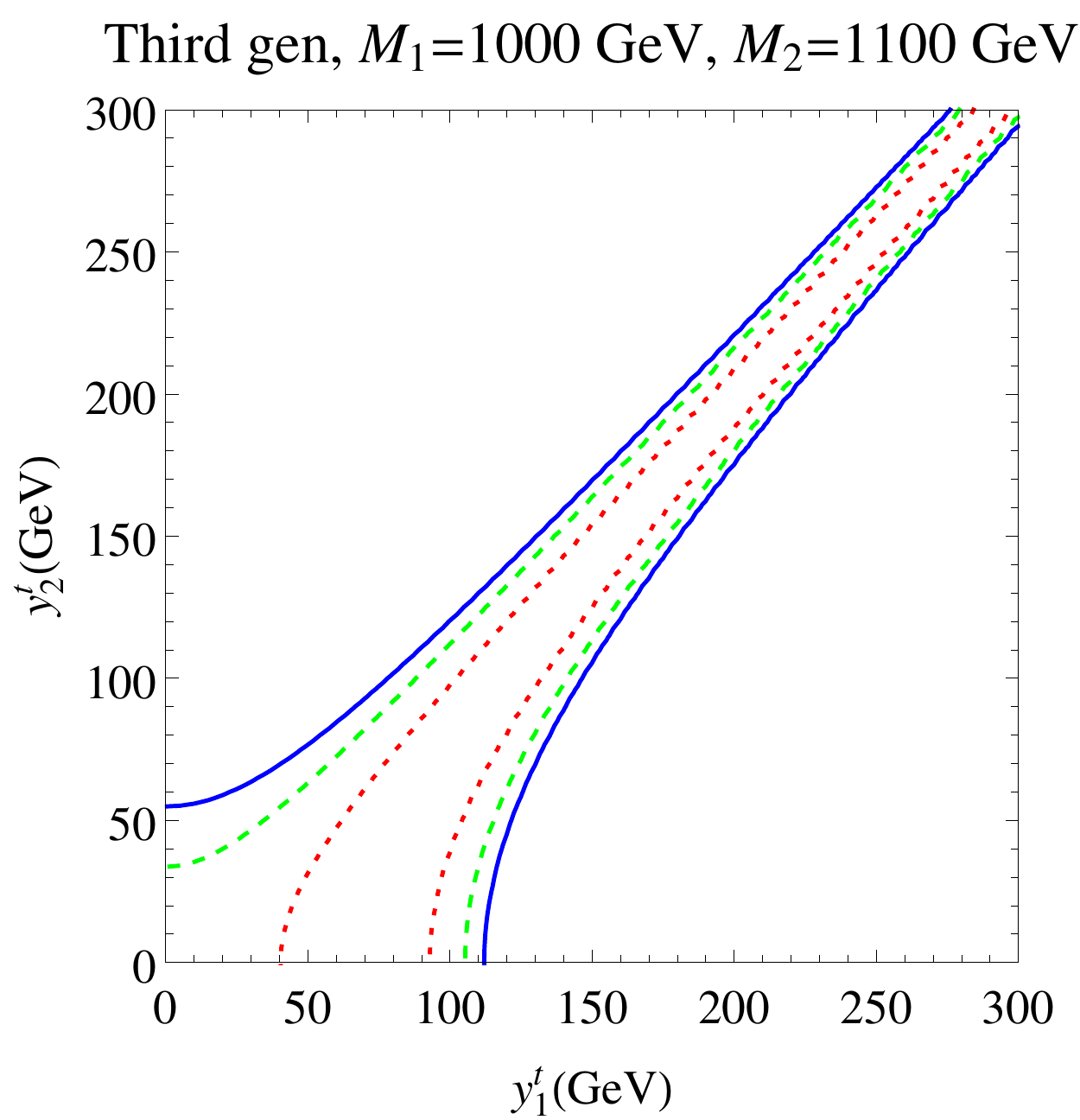,width=0.49\textwidth}\hfill
\epsfig{file=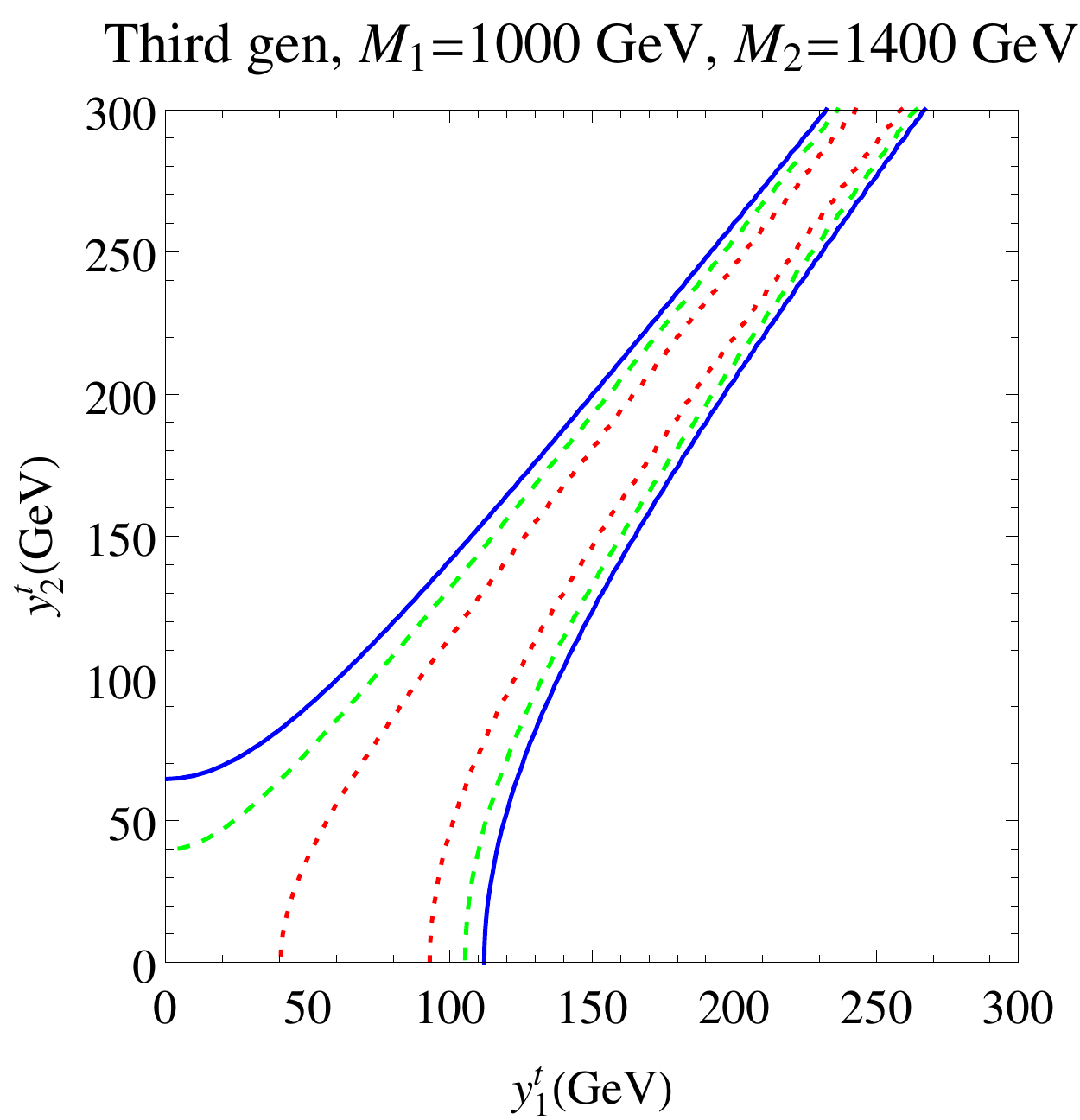,width=0.49\textwidth}
\caption{\label{fig:ewpt3rdgen}EWPT bounds (blue line is the 3$\sigma$
  bound, green dashed 2$\sigma$, red dotted 1$\sigma$, the strip
  between the lines is allowed) in the case of mixing of the two VLQ
  multiplets with the third SM quark generation.  
  }   
\end{center}
\end{figure}
%
\begin{figure}[h] 
\begin{center}
\epsfig{file=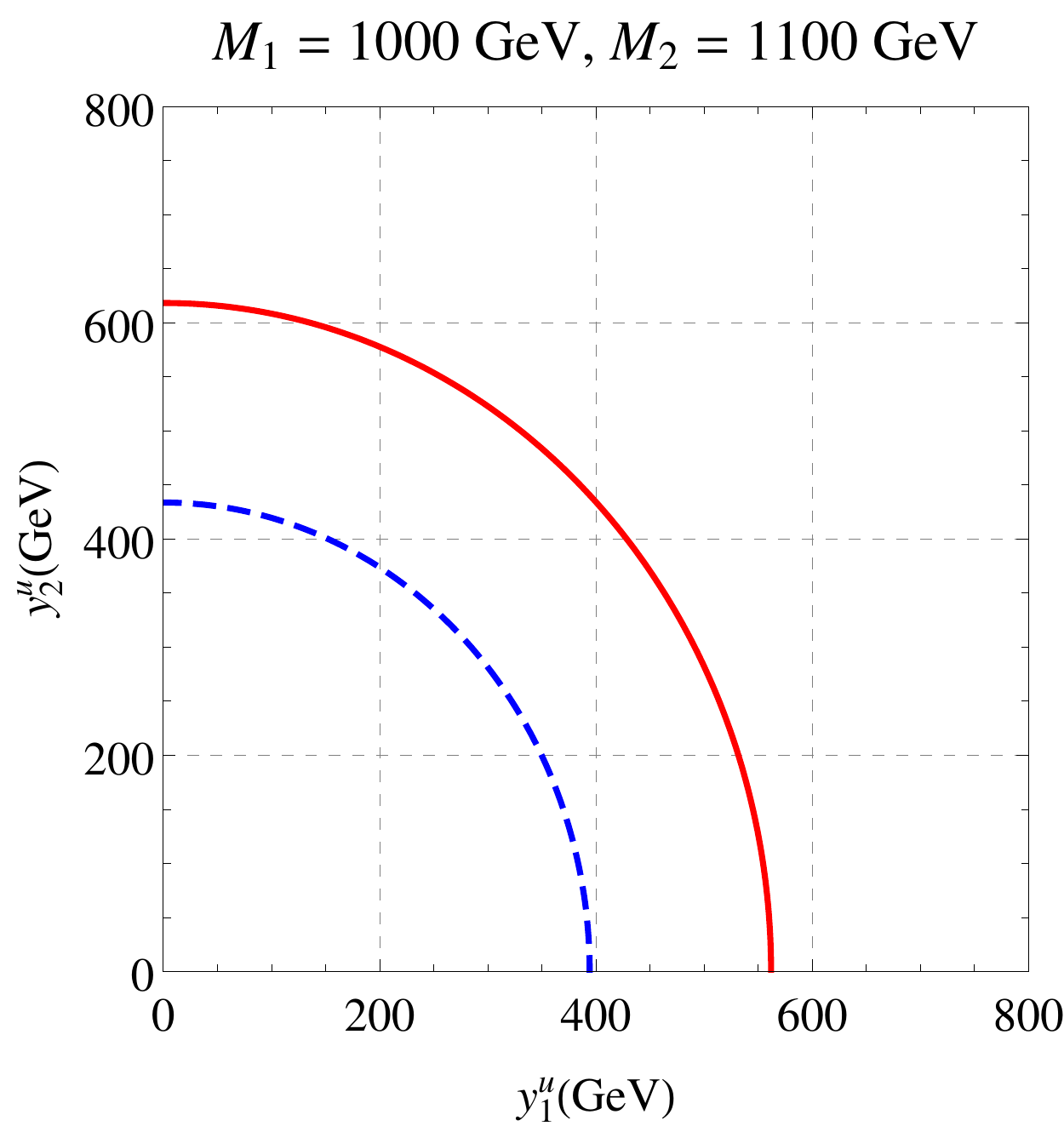,width=0.49\textwidth}\hfill
\epsfig{file=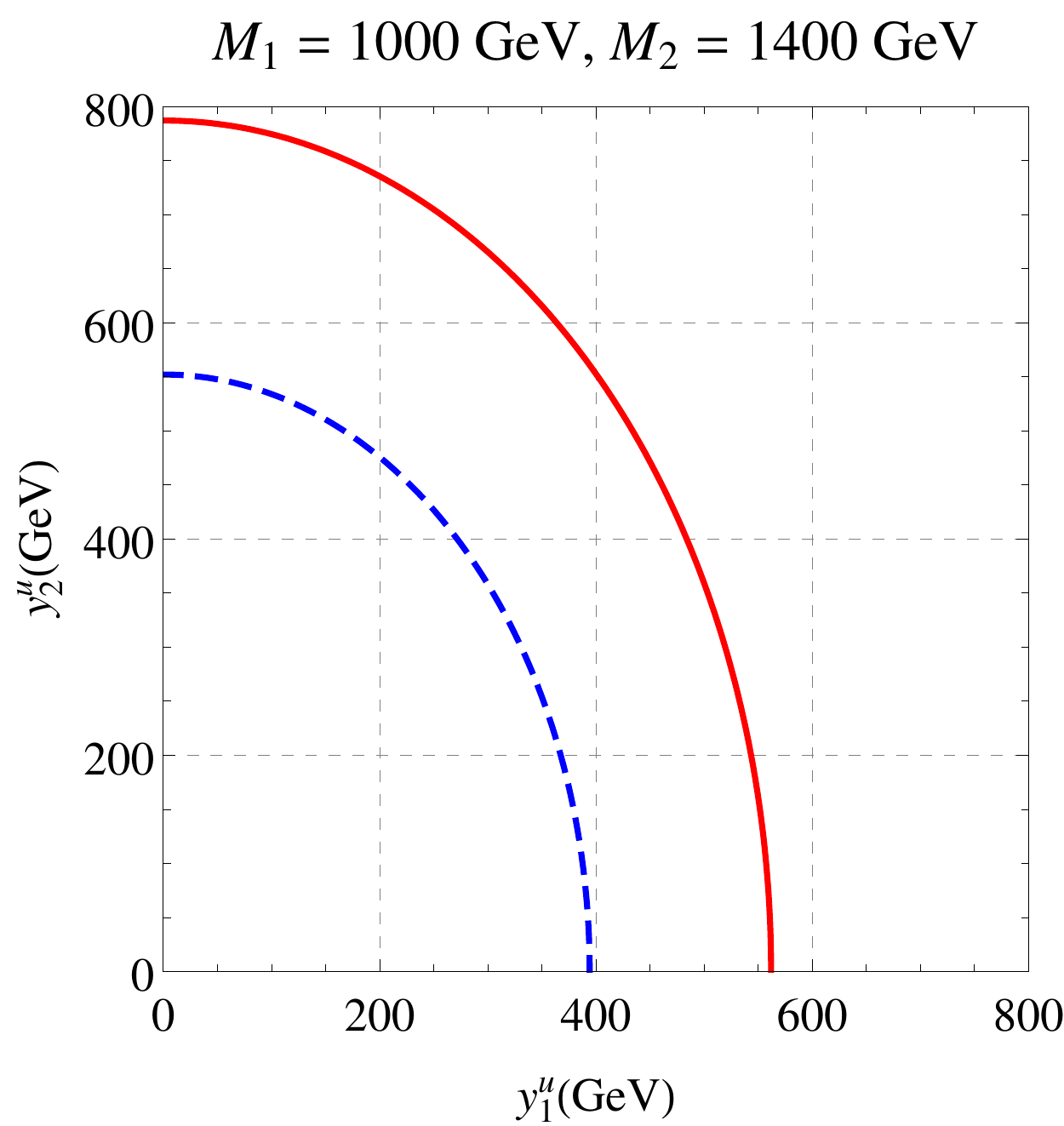,width=0.49\textwidth} 
\caption{\label{fig:higgs}First generation mixing bounds from Higgs
  couplings data, Blue dotted line is 68\% CL and Red line corresponds
  to 95\% CL. Values of the Yukawa couplings below the corresponding
  curve are allowed.} 
\end{center}
\end{figure} 
%

A combination of the numerical results we obtained are shown in
Figs.~\ref{fig:ewptlight}, \ref{fig:ewpt3rdgen} and \ref{fig:higgs}
for a selection of benchmarks. The details of the bounds we impose are
described in the following sub-sections \ref{sec:bounds:3} and
\ref{sec:EWPTHiggs}.  
The general trend is that, for VLQs that couple mainly to the first
and second SM quark families, the bounds from EWPTs (curved lines) and
tree level $Z$-couplings (excluded yellow area) tend to cover the same
parameter space. This was also remarked in
\cite{Cacciapaglia:2015ixa}, where the specific case of degenerate or
quasi degenerate multiplets was considered. For earlier discussion of
the degenerate case, we refer the reader to
Refs~\cite{Atre:2011ae,Atre:2013ap}. In the cases we cover here, with
less degenerate masses, we see that the allowed region shifts in the
parameter space of the two Yukawa couplings, while the approximate
overlap between tree and loop level bounds is conserved.  The vertical
black line gives a constraint on the Yukawa coupling coming from
direct searches for a VLQ bottom partner at the LHC Run-II (more
details in the following sub-section \ref{sec:direct}). We remark that
bounds also arise from modifications to the Higgs couplings, mainly
due to loops of VLQs to the couplings to gluons and photons. However,
such bounds (shown in Fig. \ref{fig:higgs}) are much weaker and do not
significantly affect the allowed parameter space.

The case of third generation is quite different, as there are no tree
level bounds due to our poor knowledge of the couplings of the $Z$
boson to the top quark. Furthermore, the loop contribution to the
Higgs coupling features an interesting cancellation, thus leading to
very weak constraints. The loop level EWPTs, however, give similar
constraints to the ones from light families, as shown in
Fig. \ref{fig:ewpt3rdgen}, and also features a characteristic shape
due to a cancellation that allows large values of the couplings. 

\subsection{Tree level bounds}           
\label{sec:bounds:3} 

Among the long list of processes at tree level, we consider here only
the most significant and effective one to obtain bounds on the
parameters of VLQs. Specifically, we use bounds on the modifications
to the Z couplings induced by the mixing between VLQs and SM
quarks. The couplings of VLQs to gauge bosons are given in the
appendix B of \cite{Cacciapaglia:2015ixa}. In the models under
consideration, only the mixing of top partners with up-type SM quarks
will induce this type of effects. The diagonalisation of the mass
matrix is obtained through two unitary matrices $V_L$ and $V_R$,
defined by  
\begin{equation}
M_{u} = V_L \cdot M_{u}^{diag}\cdot V_R^\dagger \,,
\end{equation}
and the mass eigenstates can be obtained by rotating the flavour
eigenstates with the same matrices: 
\begin{equation}
\left( \begin{array}{c}
u \\ c \\ t \\ t'_1 \\ t'_2
\end{array} \right)_{L/R} = V^\dagger_{L/R} \cdot \left(
\begin{array}{c} 
u^1 \\ u^2 \\ u^3 \\ U_1 \\ U_2
\end{array} \right)_{L/R}\,.
\end{equation}
The above rotations modify the couplings of SM and VLQs with the gauge
bosons, affecting in turn well measured processes, in particular
observables involving the Z boson. The expressions of couplings of
VLQs, SM quarks and the gauge bosons of the SM are provided in
Appendix~\ref{app:couplings}. The modifications to the couplings with
respect to the SM values are proportional to the $V_{L/R}^{4I}$ and
the $V_{L/R}^{5I}$ elements of the mixing matrices, and we recall that
for doublets larger mixing angles are obtained in the right-handed
sector, while the ones in the left-handed sector are suppressed by the
ratio between the SM quark mass and the VLQ masses
\cite{Buchkremer:2013bha}.  

Strong constraints on the Z coupling with first generation SM quarks
come from the weak charge measurement in atomic parity violation
experiments~\cite{Deandrea:1997wk,Patrignani:2016xqp}. The couplings
of the Z to the second generation quarks were tested in detail at
LEP~\cite{ALEPH:2005ab}:  
\begin{equation}
g_{ZL}^c = 0.3453 \pm 0.0036\,, \quad g_{ZR}^c = - 0.1580 \pm
0.0051\,, \quad \mbox{corr.} = 0.30 \,.
\end{equation}
We remark that the bounds shown in Figs \ref{fig:ewptlight} and
\ref{fig:ewpt3rdgen} are calculated at 3$\sigma$. For couplings to the
third generation, the $W_{tb}$ couplings were measured both at
TeVatron and LHC. The value of $V_{tb}$ is affected by the mixing of
the top with the VLQs in the left-handed sector:  
\begin{equation}
|V_{tb}|^2 = 1-\sum_{K=4,5} |V_L^{K3}|^2\,. 
\end{equation}
A complete list of direct measurements and lower bounds on $V_{tb}$
can be found in~\cite{Chiarelli:2013psr}. \footnote{Note that the
strong constraints from the unitarity of the CKM matrix cannot be
used, as the mixing with VLQs destroys such unitarity.} Again for
more detailed formulas we refer to \cite{Cacciapaglia:2015ixa}. 
Numerically, the bound from $V_{tb}$ are rather weak and do not
significantly affect the parameter space for heavy VLQs.

\subsection{Electroweak precision tests and Higgs bounds} 
\label{sec:EWPTHiggs}

Electroweak precision measurements, or EWPT, are a standard tool to
constrain physics beyond the SM. They can be used to constrain the
parameters of VLQs \cite{Cacciapaglia:2010vn,Chen:2017hak}, but only
under the strong hypothesis that, except for the considered
contributions, other heavy particles decouple or give negligible
contributions. Seen the level of precision in the measurement, this is
a rather strong assumption and may strongly bias the applicability of
the results to specific models. For this reason, in the following, we
will consider the bounds from EWPTs as an indication and not as a
general exclusion, contrary to the tree level bounds. The Higgs
measurements are also entering a precision era and, already at
present, give valuable information and limits on the possible
extensions of the SM. Model of VLQs are no exception and looking to
the Higgs data gives useful constraints \cite{Bizot:2015zaa}. EWPT and
Higgs couplings measurements give rather complementary bounds on the
parameters space of VLQ models.  

Bounds from EWPTs are usually given in term of the oblique parameters
S and T, as defined in Refs~\cite{Peskin:1990zt,Peskin:1991sw}.  
We have considered the following reference SM values: $m_{h, {\rm
    ref}}=125$ GeV, $m_{t,{\rm ref}}=173$ GeV and $m_{b,{\rm ref}} =
4.2$ GeV. Taking $U=0$, as it is the case in the models under
scrutiny, the experimental values for the $S$ and $T$ parameters
are~\cite{Baak:2014ora}:  
\begin{equation}
S= 0.06 \pm 0.09 \,, \qquad T= 0.10 \pm 0.07 \,,  
\label{eq:7:10}
\end{equation}
where the correlation between $S$ and $T$ in this fit is 0.91. For
more details and the complete list of formulas we refer to 
\cite{Cacciapaglia:2015ixa}.  

The EWPTs, complemented by the tree-level bounds for the light
generations, tend to favour situations in which the two Yukawa
couplings of the VLQ doublets to the SM are of similar size (see
Figures \ref{fig:ewptlight} and \ref{fig:ewpt3rdgen}), giving rise to
a funnel region that extends to large value of the Yukawas along the
diagonal. In the non-degenerate VLQ mass case, the funnel is simply
rotated away from the exact diagonal, shifting closer to the axis
relative to the heavier multiplet. This, as expected, derives from
stronger bounds on the Yukawa of the lighter multiplet. 

Concerning Higgs data, the direct measurement of the couplings to
quarks is very challenging: only very recently the observation of
production of the Higgs in association with tops has been reported by
CMS~\cite{ttH} that measured the signal strength with a $30\%$
accuracy, while the couplings to light quarks (with the exception of
the bottom) is out of reach. Thus, the only bounds come, indirectly,
from loop effects on the couplings to gluons and photons. Being
generated at loop level, they also suffer from the possible presence
of additional contributions that would thus affect the bounds in more
complete models. The combined ATLAS-CMS constraints on $\kappa_\gamma$
and $\kappa_g$ are given in Ref. \cite{exp_higgs}. The presence of new
VLQs, which enter the loops allowing the Higgs boson to couple to
photons and gluons, modifies these effective  couplings giving rise to
bounds on the parameter space of VLQs. We use therefore those combined
constraints in the following to establish bounds on the parameter
space for VLQ bi--doublets as shown in Figure \ref{fig:higgs}. These
bound put an upper limit on the funnel region which was unrestricted
by  tree-level and  EWPT data. The results of second generation mixing
with VLQs are similar to those for the first generation mixing. On the
contrary the third  generation mixing case does not allow to put any
extra constraint using the Higgs results.

\subsection{Bounds from direct searches at the LHC}
\label{sec:direct}

As we already pointed out, VLQs are widely searched for at the LHC.
Most efforts, so far, have been addressed towards VLQs that decay into
third  generation quarks and are pair produced via QCD
interactions. For a top partner, the considered final states are $W
b$, $Z t$ and $H t$. In the case of doublets, the rate into the
charged current is nearly negligible, thus leading to bounds ranging
from $1270$~GeV to $1300$~GeV from the latest CMS
results~\cite{Sirunyan:2018omb,Sirunyan:2017usq}, while
ATLAS~\cite{Aaboud:2018xuw,Aaboud:2017qpr} gives $1170$ to
$1430$~GeV. Interestingly, for CMS the stronger bound corresponds to
decays exclusively into $Z t$, while for ATLAS into $t H$. For
completeness, similar bounds can be obtained for decays into $W b$
final states~\cite{Sirunyan:2017pks,Aaboud:2017zfn}. The bounds on the
charge $-1/3$ $B$ and charge $5/3$ X, which decay uniquely into $W t$,
range between $1100$~GeV (for same sign lepton
channels)~\cite{CMS-PAS-B2G-16-019} to $1300$~GeV (for single lepton
channels)~\cite{CMS-PAS-B2G-17-008} for CMS. In the approximations
considered in the searches, those bounds do not depend on the value of
the mixing angles with the SM quarks. Searches targeting single
production channels, which are proportional to the mixing angles, are
also available within the latest dataset. CMS has published a search
for $B$ in the final state $H t$~\cite{Sirunyan:2018fjh} and for $T$
in the final state $Zt$~\cite{Sirunyan:2017ynj}, while ATLAS has a
search in $Wb$ for the 2015 dataset~\cite{ATLAS-CONF-2016-072}. Only
the search in the $Zt$ channel can, in principle, be used to set
bounds on the Yukawa-like couplings in our model. However, we have
checked that the cross sections we obtain are always smaller than the
observed bounds.

\begin{figure}[h] 
\begin{center}
\epsfig{file=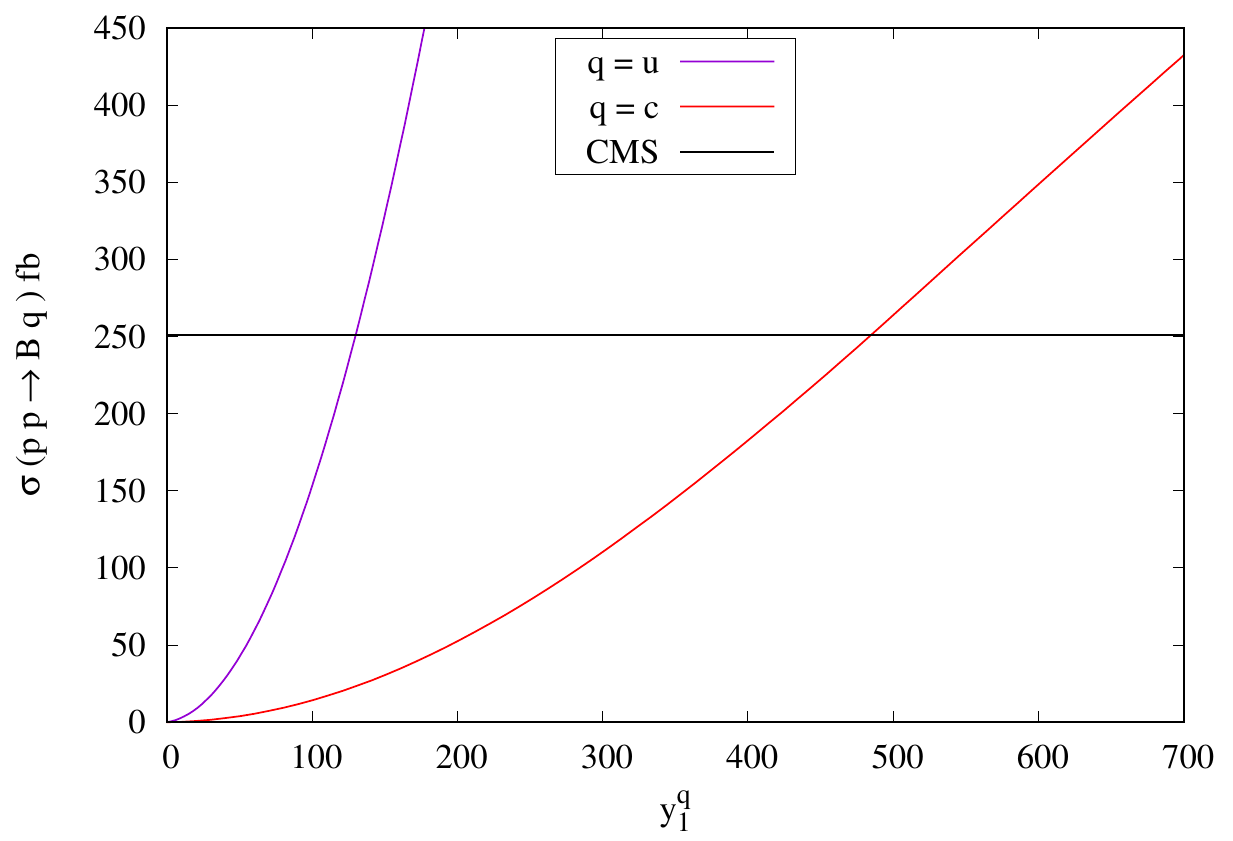,width=0.6\textwidth} 
\caption{\label{fig:direct}Cross sections for single production of a
  bottom partner $B$ in $p p \to B q$, as a function of the $y_1^q$
  (in GeV) for first and second generation mixing. The mass is fixed
  to $1000$~GeV, and the cross sections are compared to the 95\% CL
  bound from \cite{Sirunyan:2017lzl} at 8 TeV (black horizontal
  line). } 
\end{center}
\end{figure} 
%

Fewer searches also cover the case of the couplings to light
generations, and are limited to Run I data. From QCD pair
production~\cite{Sirunyan:2017lzl}, the bounds range between $430$~GeV
for exclusive decays into $H t$ to $605$ for $Z t$. Thus, our
benchmark points are well above the current exclusion. In this case,
however, single production can be very important thanks to the
couplings to valence quarks~\cite{Atre:2011ae,Buchkremer:2013bha}. 
However, interpreting
the bounds is more challenging, as they depend on the structure of the
couplings to the light quarks that enter the single production. For
the charge $2/3$ partners, in our case the dominant production is via
the couplings to the $Z$, which is however not covered in the CMS
analysis. Thus, the only bound we could directly apply to our scenario
is for the single production of a bottom-type VLQ, $B$ in the SM-like
multiplet, as cross-sections are bound by the limits for $p p \to B q$
from the CMS analysis \cite{Sirunyan:2017lzl} at 8 TeV. In turn this
provides an upper bound on the maximal value of the Yukawa couplings
for the SM-like doublet, $y_1^{u/c}$.  To extract the bound, we have
calculated the production cross section at LO, using the model
implementation described in more details in
Section~\ref{sec:cross-sections}, and compared it to the excluded
value at 95\% CL. Note that the mass of the $B$ VLQ is equal to $M_1$,
which is fixed to $1000$~GeV in our benchmarks. The result is shown in
Figure \ref{fig:direct}, where we compare the production cross section
for couplings to up (in violet) and charm (in red) quarks to the
exclusion limit at a cross section of $\sim 250$ fb. Note that we only
consider the central value here, and that an increase of the cross
section due to QCD NLO effects should be expected
\cite{Fuks:2016ftf}. Theoretical errors from scale variation are
strongly reduced at NLO. The net bounds on $y_1^{u/c}$ amount to
$y_1^u < 130$~GeV and $y_1^c < 485$~GeV, and they are shown as a black
vertical line in Figure \ref{fig:ewptlight}: the region on the left
side of the line is allowed.


\section{LHC phenomenology}
\label{sec:LHCpheno}

Having determined the allowed region in parameter space, we now
perform a phenomenological analysis of the signatures expected at the
LHC. Compared to the current search strategies, which are based on
simplified scenarios with a single VLQ, we will consider here in
detail the interplay between the two VLQ doublets with hypercharges
$Y=1/6$ and $Y=7/6$. We will show that peculiar patterns in the decay
rates can be observed, as well as new production channels.  

Among the key properties of this scenario is the presence of two top
partners that mix and have different masses and decay patterns. One
feature common to all top partners coming from doublets is that the
decay via charged currents, {\it i.e.} a $W^\pm$ boson, are very
suppressed, thus searches based on this decay channel (which give the
strongest bounds) will be ineffective. As we will see, peculiar decay
patterns may be used to effectively tag this kind of scenario. 

\subsection{Masses and branching ratios}

\begin{figure}[tb]
\begin{center} 
\epsfig{file=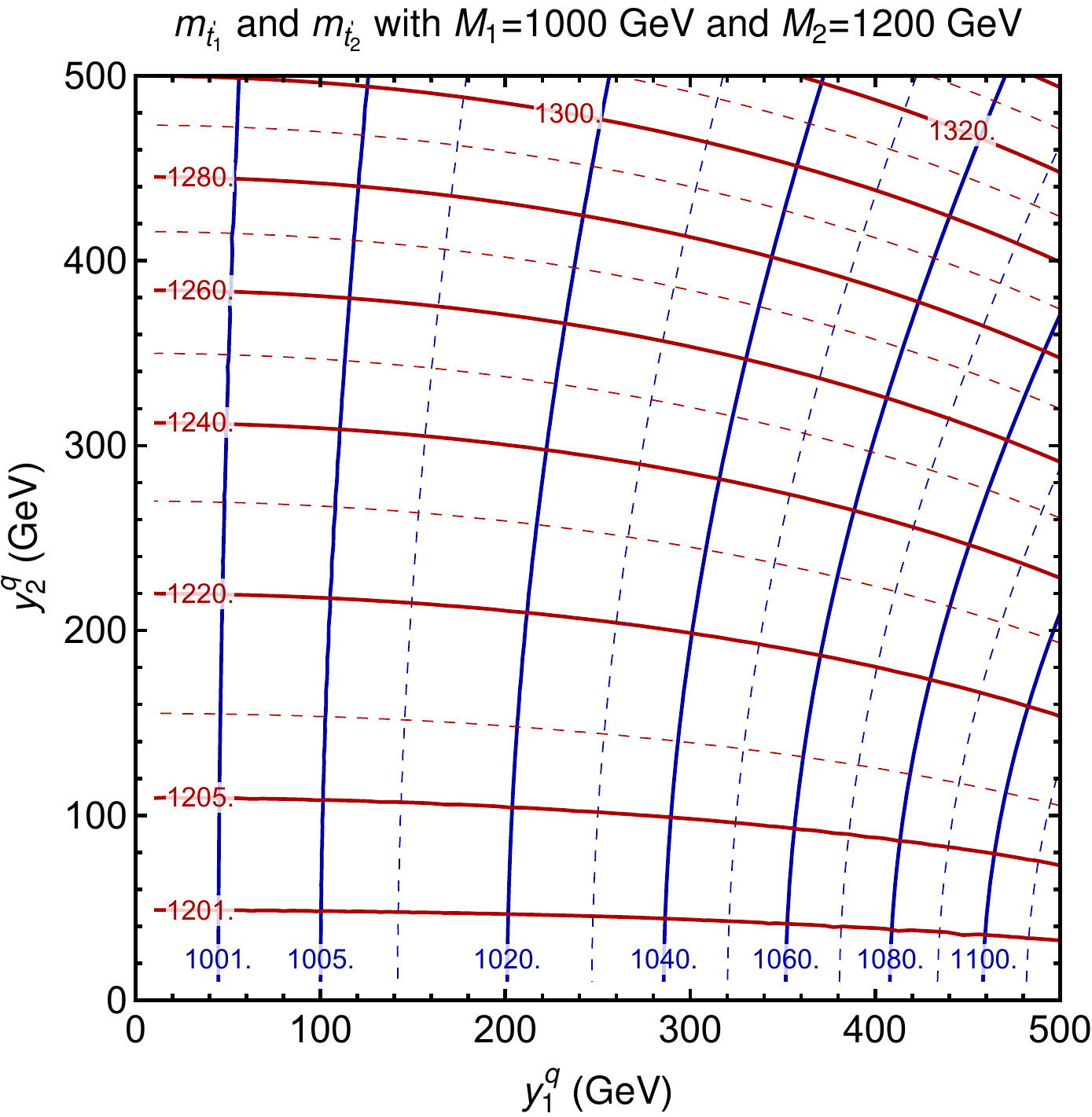,width=0.48\textwidth}
\caption{\label{fig:masses} Masses in GeV of $t'_1$ (blue contours) and $t'_2$ (red contours) mixing with the light quarks, for the benchmark $M_1=1000$~GeV and $M_2=1200$~GeV. The contours are shown at intervals of 10 GeV unless specified. Results for mixing with the top quark are numerically almost identical.} 
\end{center} 
\end{figure}

The analytical expressions of the masses and branching ratios (BRs)
are reported in Appendices \ref{app:masses} and \ref{app:BRs}
respectively.  We recall that the values of the masses for $t'_1$ and
$t'_2$ are not constant but depend on the values of the two Yukawas,
as shown in Fig.~\ref{fig:masses}. We show results for the light
quarks and for the benchmark masses $M_1=1000$~GeV and
$M_2=1200$~GeV. For mixing with the top, the results are qualitatively
similar and quantitatively very close too, as the VLQ masses are
already constrained to be much heavier that the top, as discussed in
the previous section. On the other hand, the bottom-partner $B$ and
exotic charged $X^{5/3}$ have masses fixed, respectively, to $M_1$ and
$M_2$, and BRs of 100\% into  $B \to W^- u/c/t$ and $X^{5/3} \to W^+
u/c/t$. 

For the branching ratios, in this section we present sample numerical
results for the intermediate benchmark scenario with $M_1=1000$ GeV
and $M_2=1200$ GeV, as the results for the other two cases as well as
for heavier masses are qualitatively similar.  

\begin{figure}[tb]
\begin{center} 
\epsfig{file=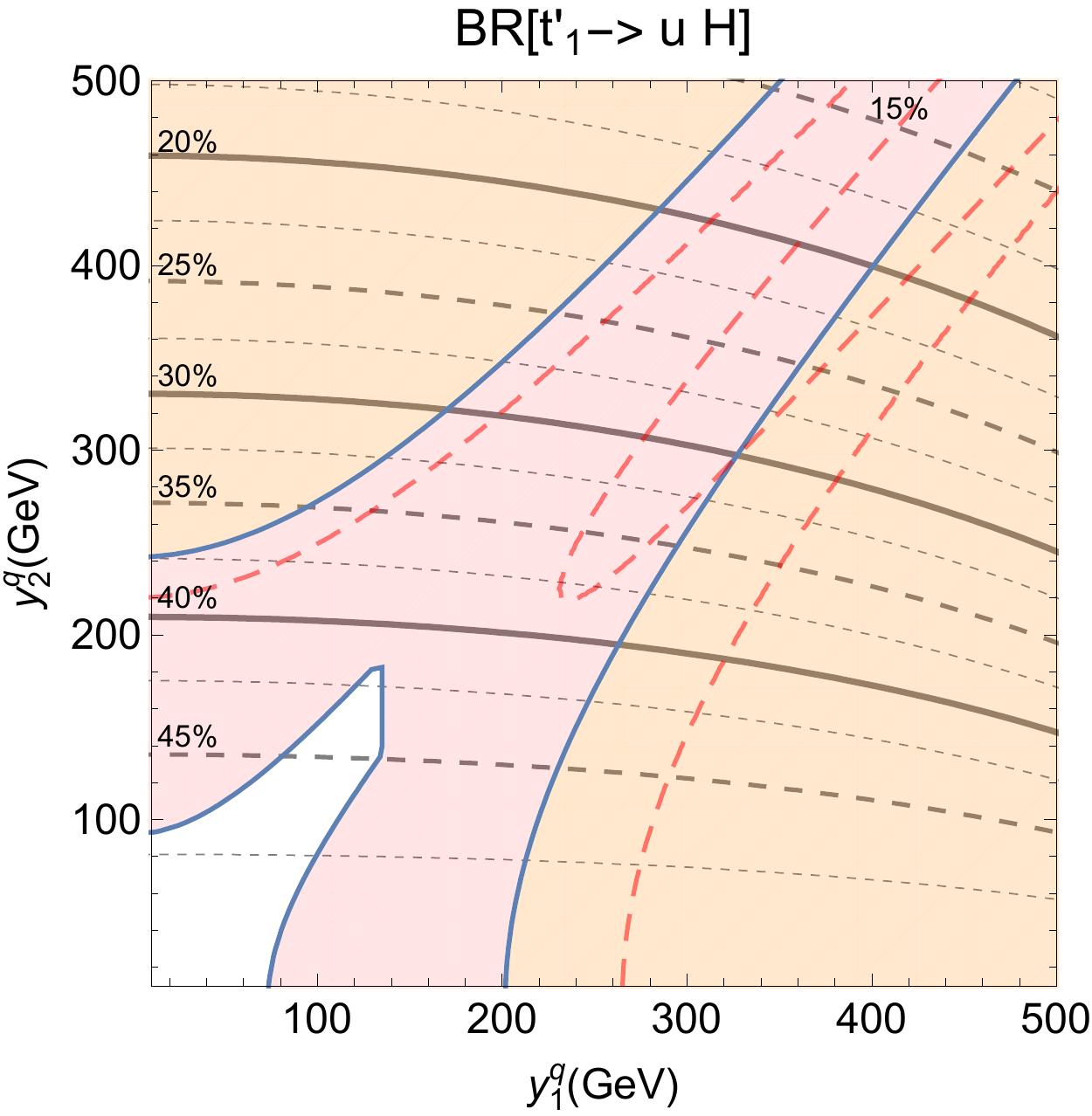,width=0.47\textwidth}\hfill
\epsfig{file=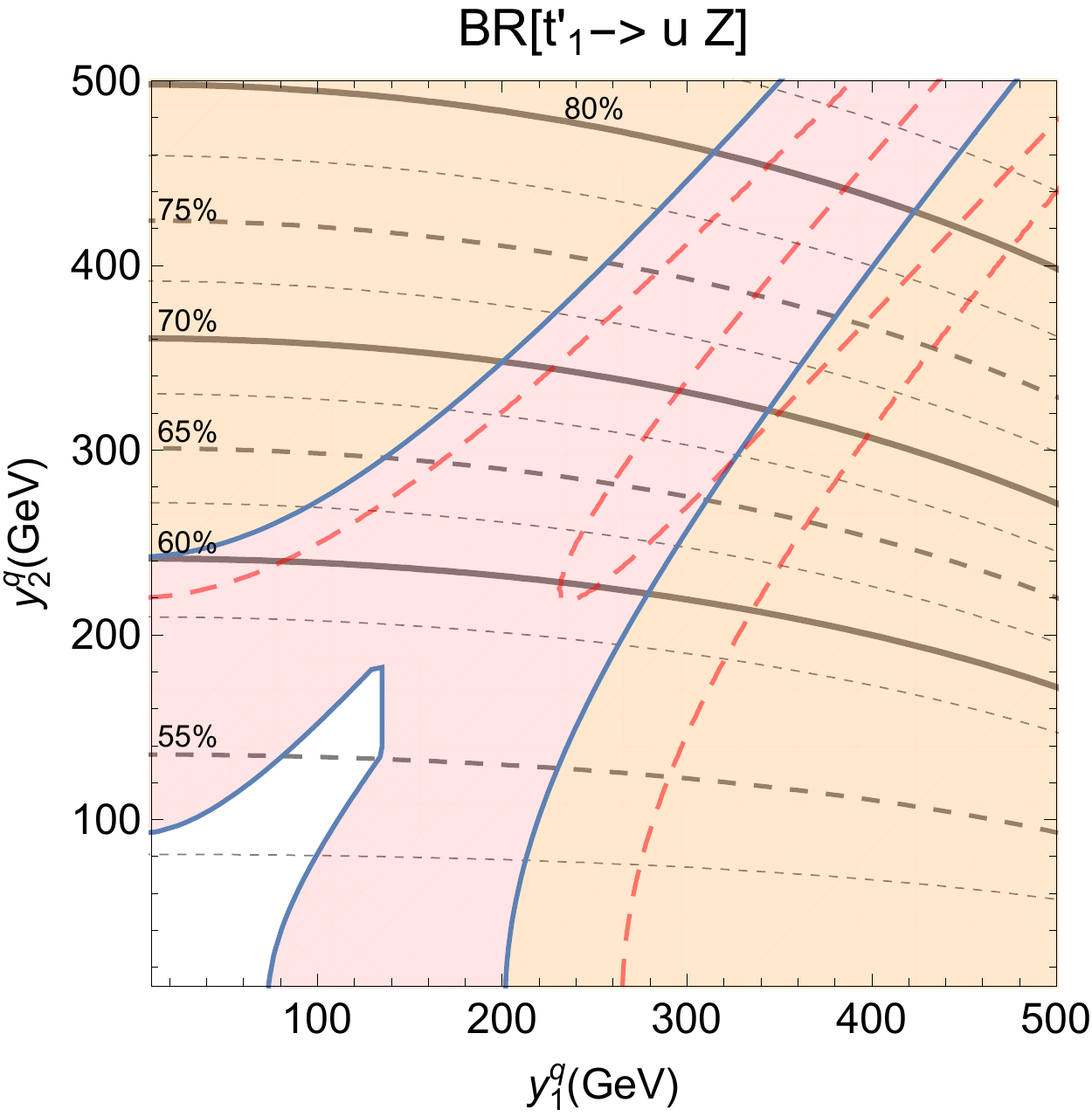,width=0.47\textwidth}
\caption{\label{fig:brtp1_M1012} Branching ratios of $t_1^\prime$
  mixing with the up/charm quark . The contours are show for values of
  the BR spaced by steps of $2.5\%$. For the light quarks, the orange
  region is excluded for the charm, while the orange plus pink areas
  are excluded for the up. The dashed line indicate the region
  excluded by EWPTs for mixing to the charm.  } 
\end{center} 
\end{figure}

We start from the lighter top partner, $t_1^\prime$. In
Fig.~\ref{fig:brtp1_M1012} we show contours of the BRs of a
$t_1^\prime$ that mixes with the up quark. The contours are shown in
the plane identified by the two Yukawa couplings. Results for mixing
to the charm are nearly identical (differences only depend on the mass
of the charm, which is much smaller that the VLQ masses), so we
superimpose on the same plot the regions excluded by tree level
constraints for the two cases: orange for the charm, with the  pink
area additionally excluded for the up. The orange line marks the
additional portion of parameter space that would be excluded at
3$\sigma$ by the loop-level EWPTs, in absence of additional
contribution from New Physics and for mixing to the charm (for the up,
the tree level bounds are always dominant). We notice that the charged
current is absent, and that the decay rates are mostly sensitive to
the value of the Yukawa for the second multiplet. For small values of
$y_2^q$, the rates are almost equal between $Z$ and Higgs, while at
large values the $Z$ tends to dominate. The analogous BRs for mixing
of $t^\prime_1$ to the top  are numerically very similar, due to the
smallness of the top mass compared to the VLQ ones while , however,
the excluded region is different (recall the absence of tree-level
constraints). 

\begin{figure}[tb]
\begin{center} 
\epsfig{file=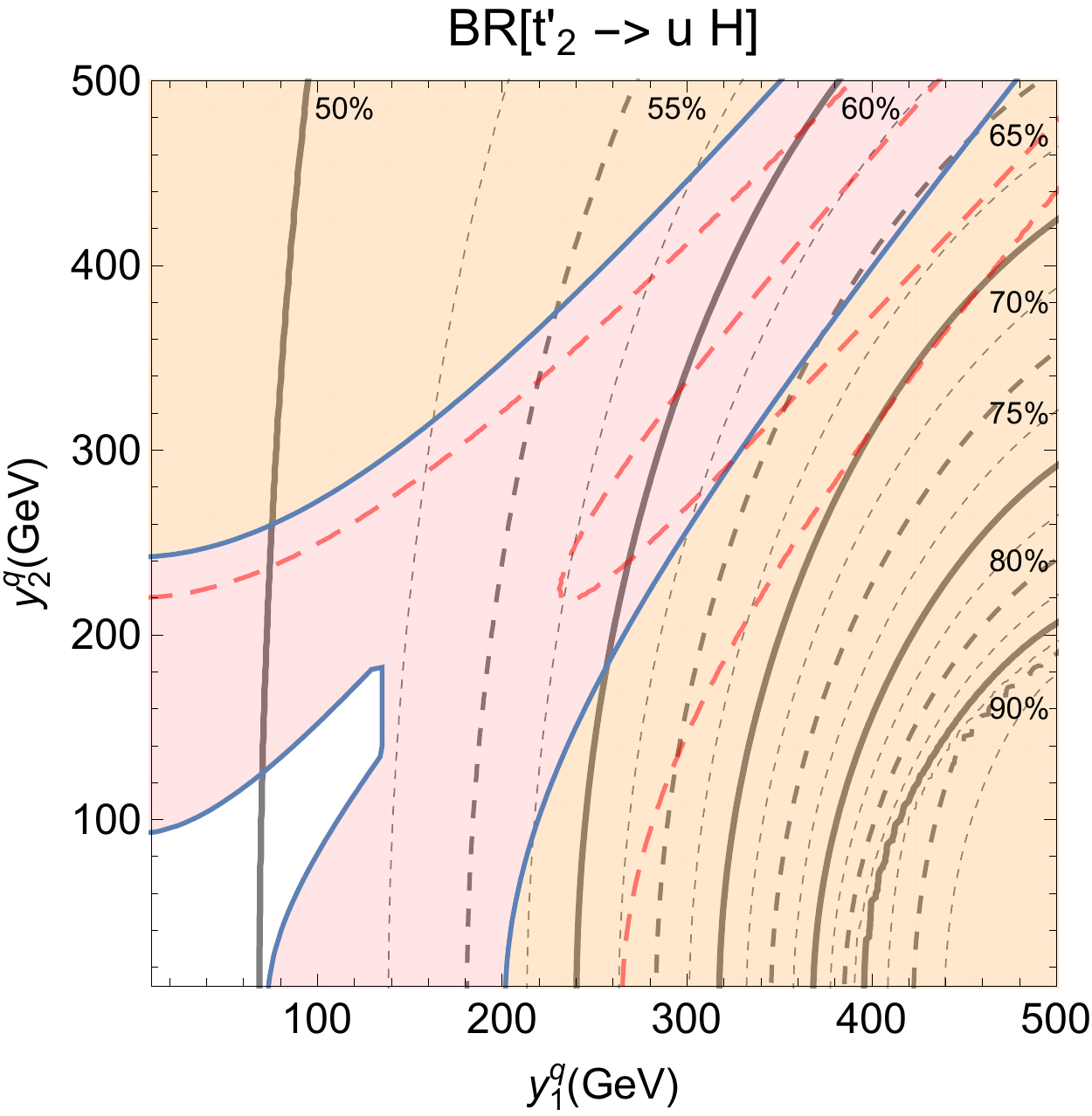,width=0.47\textwidth}\hfill
\epsfig{file=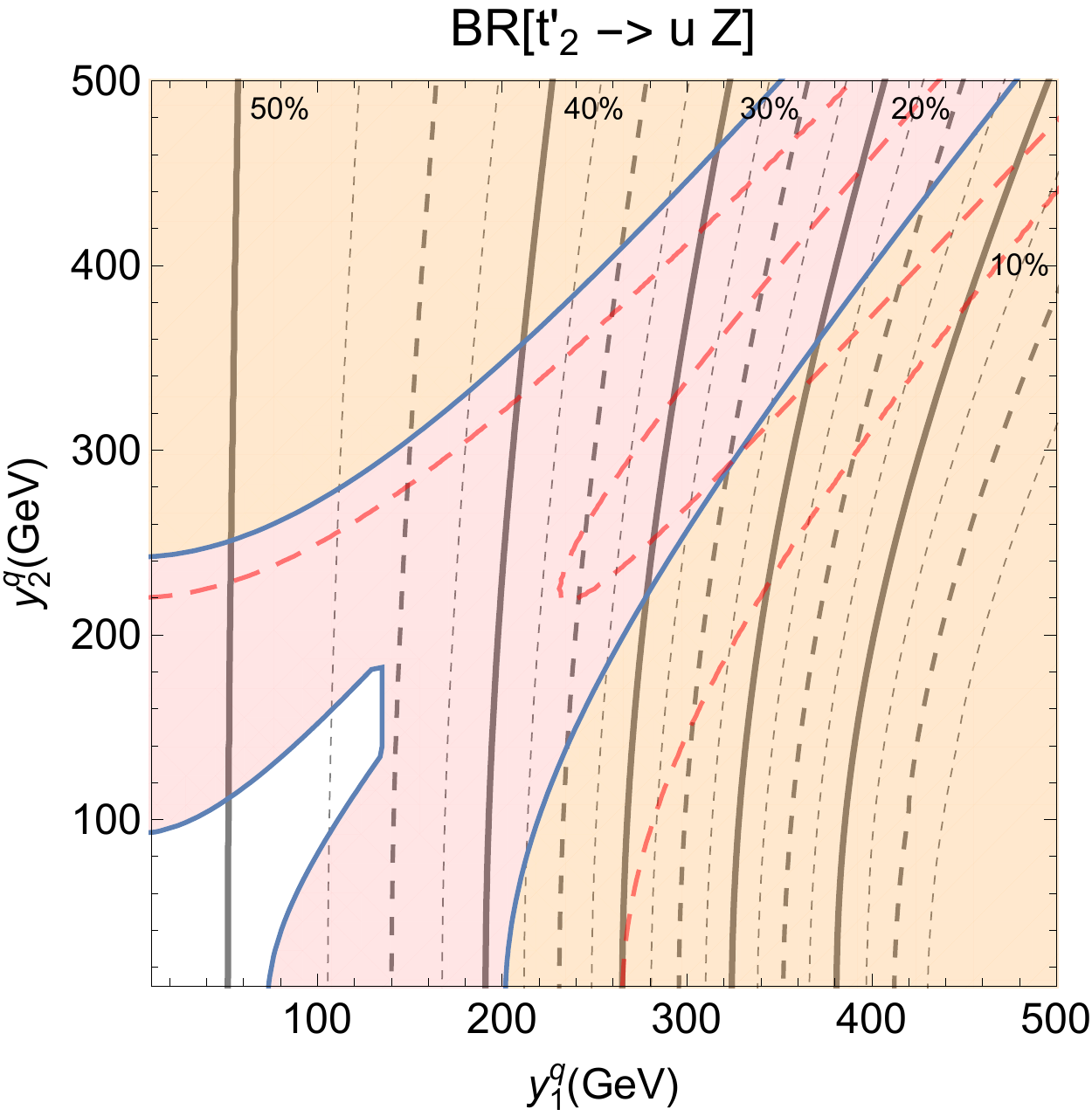,width=0.47\textwidth}\\[8pt]
\epsfig{file=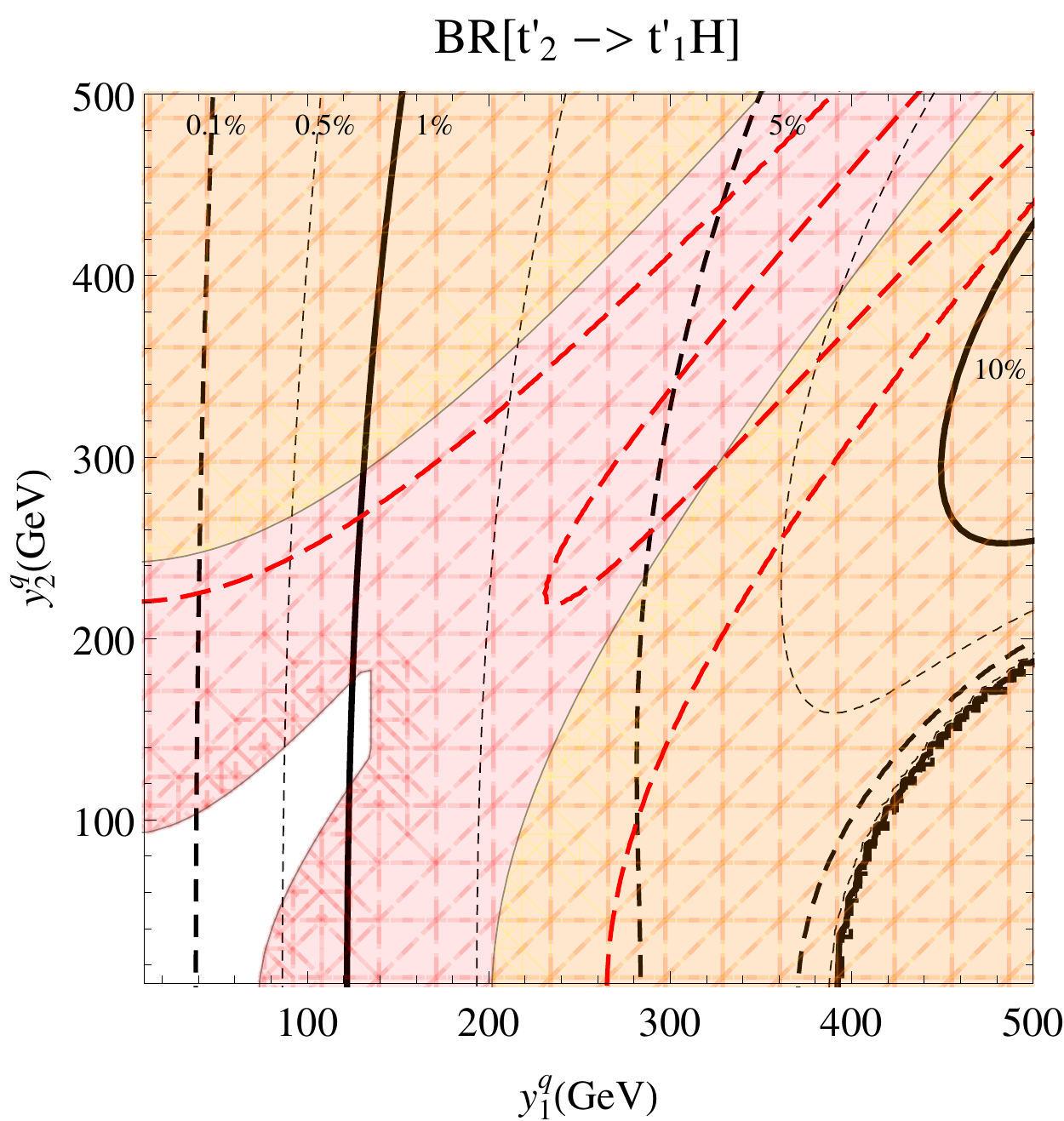,width=0.47\textwidth}\hfill 
\epsfig{file=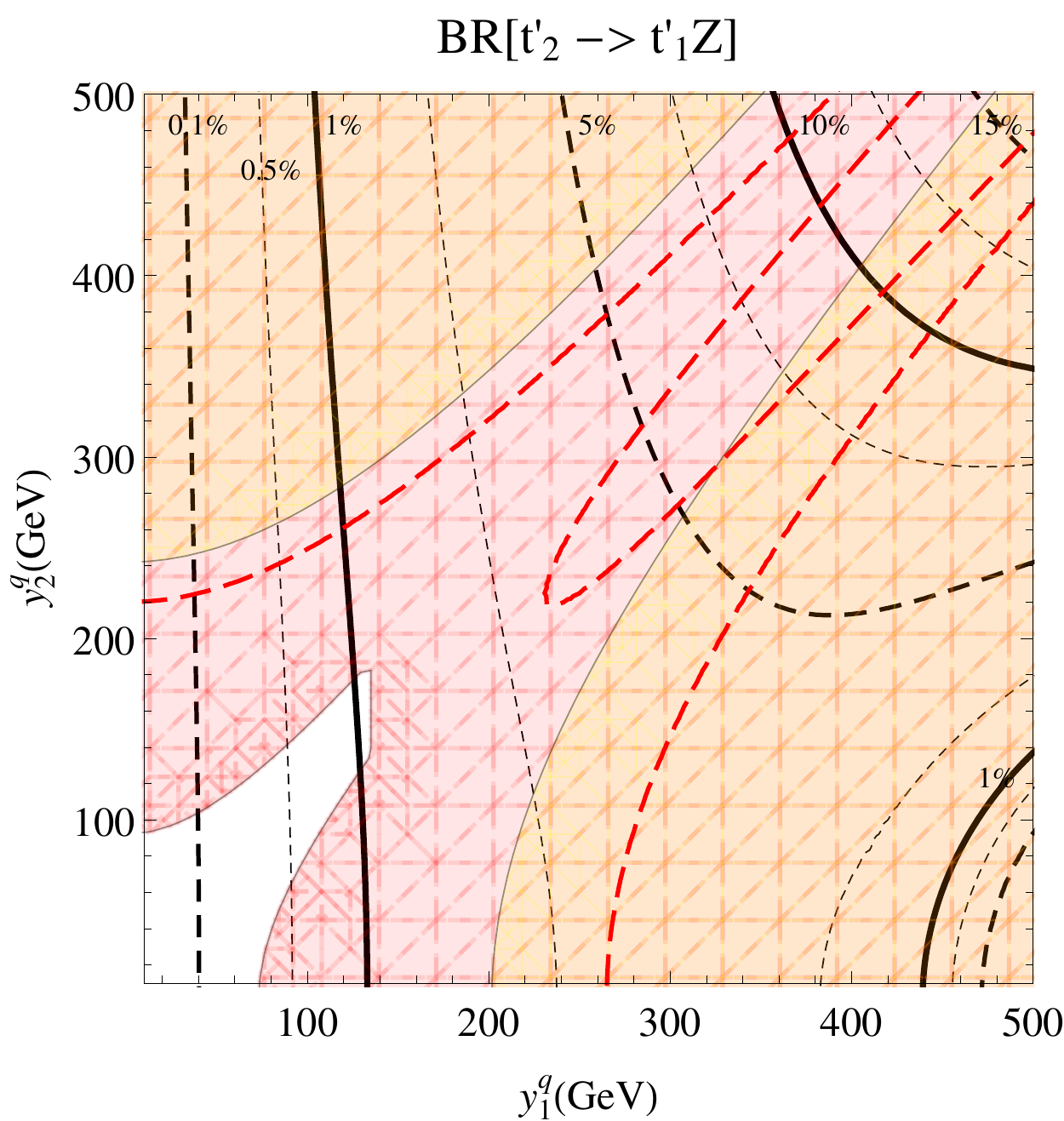,width=0.47\textwidth} 
\caption{\label{fig:brtp2_M1012_1stgen}Branching ratios of
  $t_2^\prime$ mixing with the up/charm quark. The contours are show
  for values of the BR spaced by steps of $2.5\%$ (unless
  specified). The orange region is excluded for the charm, while the
  orange plus pink areas are excluded for the up. The dashed line
  indicate the region excluded by EWPTs for mixing to the charm.}  
\end{center} 
\end{figure}

For the heavier $t_2^\prime$, we show the BRs in
Figs~\ref{fig:brtp2_M1012_1stgen} for couplings to light
generations. We note the same pattern in the balance between the $Z$
and Higgs final states, but with inverted roles: it is the BR into the
Higgs that dominates, in this case, for large values of the Yukawa
with the first doublet, $y_1^q$. In addition, decays into the lighter
VLQ $t^\prime_1$ are also allowed, but with very small rates that only
increase above the few percent for large Yukawa couplings. 

It is useful to remark that, for mixing to the up quark, the allowed
parameter region is very small, thus the values of the BRs are
constrained to almost fixed values. For both $t^\prime_1$ and
$t^\prime_2$, the rates into $u Z$ and  $u H$ are close to $50\%$,
while decays $t^\prime_2 \to t^\prime_1 Z/H$ are always bound to be
below $1\%$. 

\clearpage
\subsection{Cross-sections} \label{sec:cross-sections}

The production cross-sections at the LHC also show distinctive
patterns. For the calculation, we have used a modified version of the
{\sc Feynrules}~\cite{Alloul:2013bka} VLQ model files provided in
Ref.~\cite{Fuks:2016ftf}. A modification is necessary for including
couplings between VLQs from different multiplets and SM gauge and
Higgs bosons\footnote{The modified FeynRules file is available here:
  \url{http://deandrea.home.cern.ch/deandrea/VLQ_v4.fr}}. Such
modifications allow an estimation of processes where VLQs of different
multiplets are produced in association, as $p p \to t_1^\prime
t_2^\prime$. We have used Madgraph5 version 2.6.1 \cite{Alwall:2011uj}
for the estimation of cross-sections at LO in QCD, using the NN23LO1
parton distribution functions for the proton.   

\begin{figure}[htb]
\epsfig{file=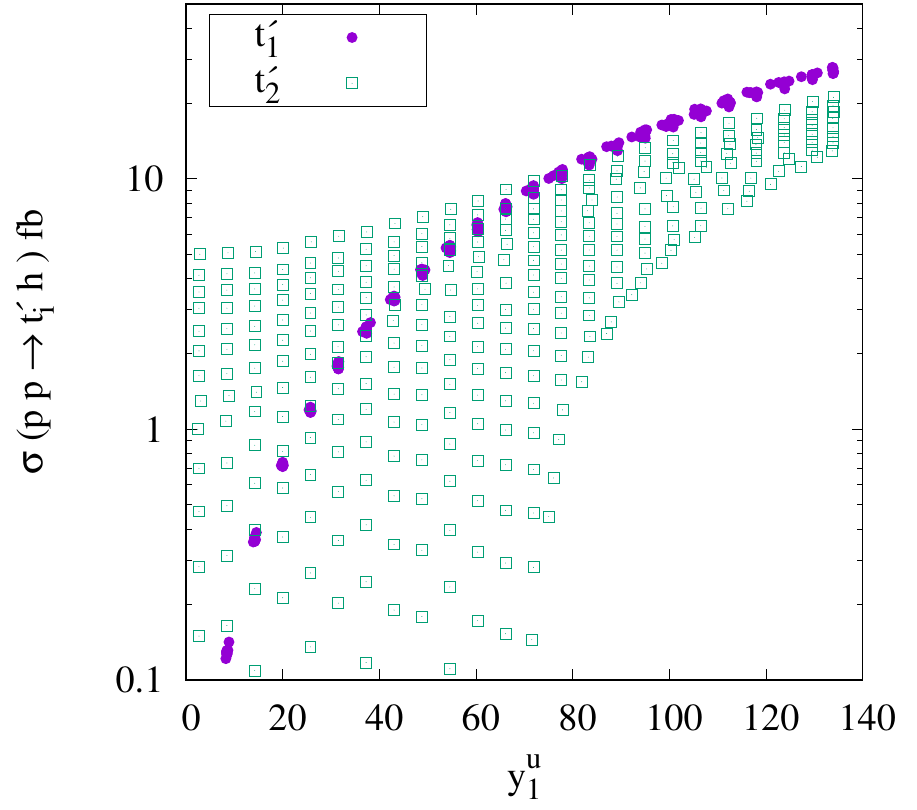,width=0.5\textwidth}
\epsfig{file=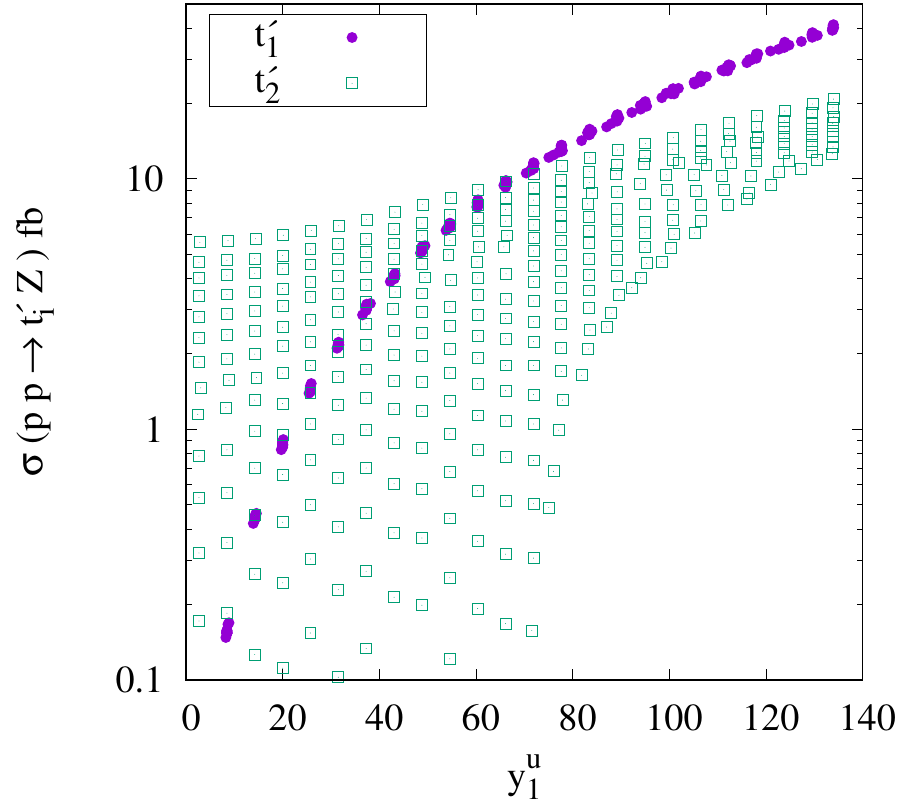,width=0.5\textwidth}
\\
\epsfig{file=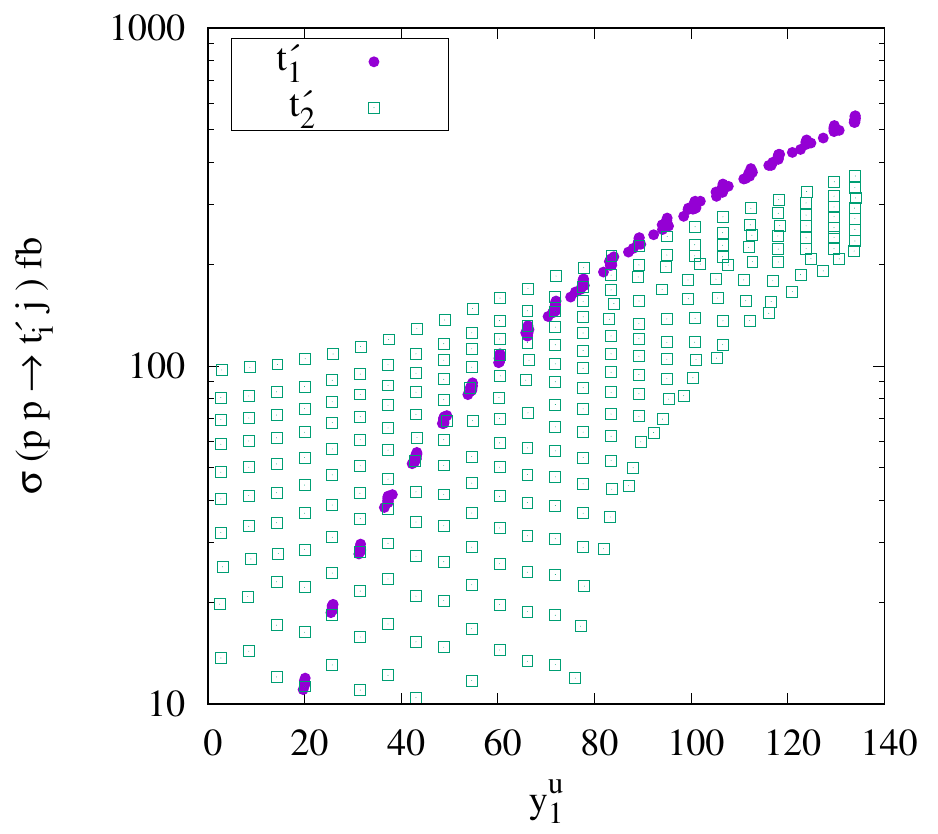,width=0.5\textwidth}
~~\epsfig{file=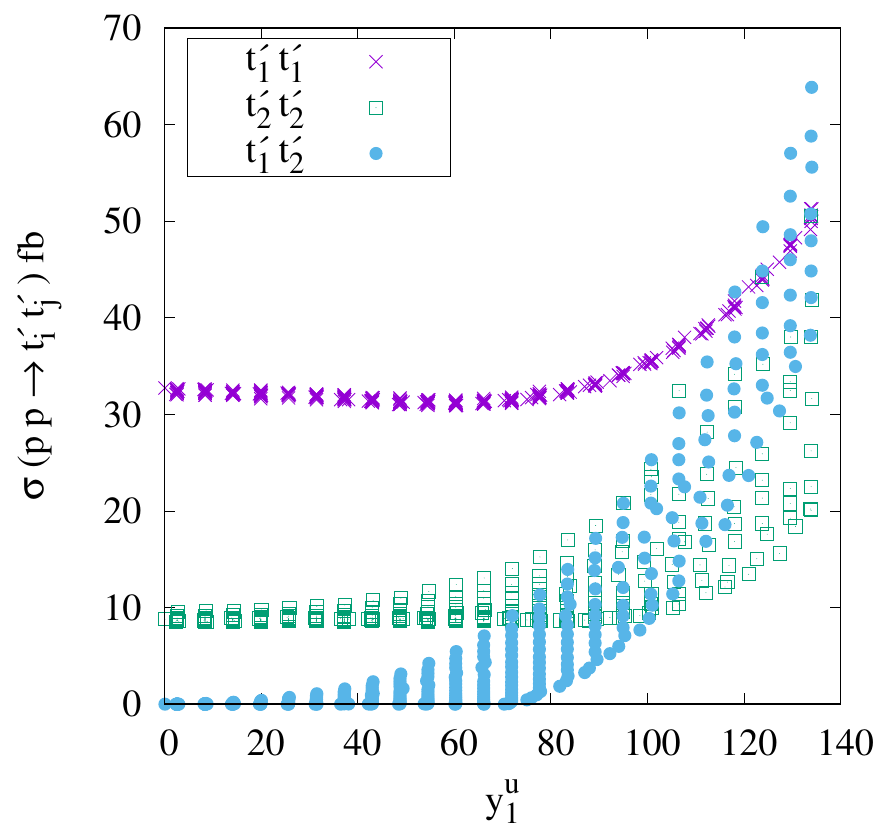,width=0.48\textwidth} 
\caption{\label{fig:LHC_13TeV}Scatter plot of production
  cross-sections at LHC@13TeV, scanning over the Yukawa coupling of
  $t_2^\prime$, for the production processes (from top left
  clockwise): $t_i^\prime+h$, $t_i^\prime+Z$, $t_i^\prime t_j^\prime$
  and $t_i^\prime+\text{jet}$ with $(i,j = 1,2)$ and with $M_1=1000$
  GeV and $M_2=1200$ GeV. The cross-section include also the
  production of the anti-particle states $\overline
  t_{i,j}^\prime$.}
\end{figure}

We computed the production cross-sections at 13 TeV for the
production of the charge $2/3$ VLQs $t_{1,2}^\prime$ in the parameter space
allowed by precision, low energy and LHC@8TeV constraints, determined
in Section \ref{sec:ewbounds}. We also focus on mixing to the up quark,
which allows for sizeable single production rates thanks to the couplings
to a valence quark in the proton. We chose, as representative
benchmark, the set of input parameter $M_1 = 1000$ GeV and $M_2 = 1200$
GeV and  scanned over the allowed values of the Yukawa couplings
in the $y_1^u$ - $y_2^u$ plane. Specifically, we have considered
processes of single production of $t_1^\prime$ and $t_2^\prime$ in
association with SM objects, $p p \to t_{1,2}^\prime+\{h,Z,j\}$, and
pair production of top partners of same or different kind, $p p \to
t_i^\prime t_j^\prime$ with $i,j=1,2$, therefore including both QCD-
and EW-strength couplings. In all cases we have considered both the
production of particle and anti-particle states. Our results are
summarised in Fig.~\ref{fig:LHC_13TeV}.  

A number of conclusions can be derived:
\begin{itemize}
\item In the allowed region of parameter space, it is always possible
  to obtain configurations in which the production cross-section of
  the heavier VLQ ($t_2^\prime$) is comparable or even larger than the
  cross-section for the lighter VLQ ($t_1^\prime$). For single
  production channels, this switch happens for values of $y^u_1$
  smaller than $60\div 80$ GeV, while for pair production channel the
  production of $t_2^\prime$ is comparable to $t_1^\prime$ for values
  of $y_1^u$ around the upper allowed limit. From a phenomenological
  point of view, this result can be very interesting because the decay
  patterns of the heavier top-partner are different from the ones
  usually considered in experimental searches. This includes the
  possibility of chain-decays to the lighter $t_1^\prime$, thus
  opening new channels for experimental exploration.  

\item The cross-section for production of a pair of VLQs of same kind
  exhibits an interesting pattern that indicates the dominance of EW
  production mechanism for large values of Yukawa couplings. In fact,
  QCD production only depends on the mass, which depends only mildly
  on the Yukawas. From the bottom-right plot in
  Fig.\ref{fig:LHC_13TeV}, however, we see a marked increase in the
  cross-section of both VLQs for $y_1^u > 60 \div 80$~GeV. In
  addition, the pure electroweak production of the two VLQs together,
  $t_1^\prime t_2^\prime$,  becomes sizeable in the same parameter
  region, and it even dominates over pair production for the largest
  allowed values of the Yukawa couplings. This may be extremely
  relevant for phenomenological analyses as the kinematics of
  processes of production of a pair of VLQ with different masses will
  be different from the one usually considered in experimental
  searches where the same VLQ is produced in pairs only through
  QCD-driven processes. A similar effect was noted in
  Ref.~\cite{Cacciapaglia:2009cu} for VLQs in the context of Little
  Higgs models with T-parity and, more recently, the same phenomena
  was noted in \cite{Brooijmans:2018xbu,Fuks:2016ftf}.  
\end{itemize}

The results we highlighted show novel channels that deserve a
thoroughly investigation, as they may give rise to detectable
characteristic signatures at the LHC. Furthermore, an analysis at NLO
in QCD~\cite{Fuks:2016ftf,Cacciapaglia:2017gzh} is needed to go beyond
a simple cross-section calculation, together with the addition of
detector and reconstruction effects.  

\section{Conclusions}          
\label{sec:concl}
We have considered VLQs in a more general framework than the usual
simplified models, namely we study the presence of two doublets with
general mixing structure with the SM quark generations. This template,
inspired by situations which are typically present in various NP
models, shows that present bounds in the general case are weaker than
those assuming a single VLQ multiplet and coupling only to the third
SM quark generation. Moreover we focused on the two
``top-partner-type'' heavy VLQs present in the case of the two studied
multiplets. Due to their peculiar mixing patterns with the SM quarks,
they feature production and decay channels that are usually not
considered in experimental searches. In particular, we remark areas in
the parameter space for sizeable Yukawa couplings where the single
production of the heavier partner dominates, thus leading to cascade
decays. Furthermore, in the same parameter region, production of the
two mass eigenstates in association can dominate over QCD and EW pair
production. These new features deserve to be included within the
exploration programs for NP at the LHC, thus allowing to test these situations in detail.



\section*{Acknowledgments}
AD is partially supported by the Institut Universitaire de France. AD and
GC also acknowledge partial support from the Labex-LIO (Lyon Institute
of Origins) under grant ANR-10-LABX-66, FRAMA (FR3127, F\'ed\'eration
de Recherche ``Andr\'e Marie Amp\`ere''). AD, GC and NG would like to
acknowledge the support of the CNRS LIA (Laboratoire International
Associ\'e) THEP (Theoretical High Energy Physics) and the INFRE-HEPNET
(IndoFrench Network on High Energy Physics) of CEFIPRA/IFCPAR
(Indo-French Centre for the Promotion of Advanced Research). The work of 
AD, GC and NG was also supported by CEFIPRA/IFCPAR grant number 5904-C. The
research of YO is supported in part by JSPS KAKENHI Grant Number
JP15K05066. This work is also supported in part by the TYL-FJPPL
program. The work of DH is supported by grant number NSFC-11422544.

\appendix
\section{Appendix}
\subsection{Multiplets}
\label{app:multiplets}
The doublets we consider allow to have mixing with the SM and one
state with same quantum numbers in the two multiplets, and in the main
text we focused  on the situation in which each doublet contains a
top-partner. However it is also possible to have two bottom partners,
giving rise to the following two cases of either two $T'$ VLQs or two
$B'$ VLQs. 

\subsubsection*{Case 1) Doublet $Y=1/6$ and Doublet $Y=7/6$}
In the case of two VLQ doublets, due to their quantum numbers, they
couple to the right-handed SM quarks: 
\begin{eqnarray}
\mathcal{L}_{V-SM} &=& 
    - \lambda_{1}^k\, \bar{\psi}_{1L} \tilde{H} u_R^k -
    \lambda_{1d}^k\, \bar{\psi}_{1L} H d_R^k - \lambda_2^k \,
    \bar{\psi}_{2L} H u_R^k + h.c. \,,  \label{eq:LV-SM}
\end{eqnarray}
where the VLQ fermions are $\psi_1 =({\bf 2},\frac{1}{6})=\left(U_1 \ D_1\right)^T$ and $\psi_2 =
({\bf 2},\frac{7}{6}) = \left(X_2^{5/3} \ U_2\right)^T$. 
No Yukawa coupling between the two VLQ multiplet is allowed, therefore one can use two free phases to remove one phase in 
$\lambda_1^k$ and one in $\lambda_2^k$ (therefore only four new phases are present). 
The mass Lagrangian is:
\begin{eqnarray}
\mathcal{L}_{\rm mass} &=& 
  - y_{1u}^k \bar{U}_{1L} u_R^k -  y_{1d}^k \bar{D}_{1L} d_R^k - y_2^k
  U_{2L} u^k_R                                   \nonumber \\ 
&&  - M_1 \, \bar{U}_{1L} U_{1R} - M_1 \, \bar{D}_{1L} D_{1R} -  M_2
\, \bar{U}_{2L} U_{2R} -  M_2 \, \bar{X}^{5/3}_{2L} X^{5/3}_{2R} + h.c. \,, 
\end{eqnarray}
and mass matrices are:
\begin{equation}
M_u = \left( \begin{array}{c c c}
\left(\tilde{m}^{up}\right)_{3\times 3} &  0_{3 \times 1} &
0_{3\times1} \\ 
 (y_{1u}^k)_{1\times 3} & M_1 &  0 \\
(y_2^k)_{1 \times 3}  & 0 & M_2
\end{array} \right), \quad
M_d = \left( \begin{array}{c c }
\left(\tilde{m}^{down}\right)_{3\times 3} & 0_{3\times 1}  \\
(y_{1d}^k)_{1\times 3} & M_1 \\
\end{array} \right), \quad 
M_{X^{5/3}} = M_2 \,.    
\end{equation}

\subsubsection*{Case 2) Doublet $Y=1/6$ and Doublet  $Y=-5/6$} 

We have not considered in the main text the case of two bottom
partners as it requires a different study, implying also the use of
quite different bounds. We give it here for completeness and future
reference for further studies. In this case the the Lagrangian
$\mathcal{L}_{V-SM}$ differs from the one of the previous case due to
the different weak hypercharge of the second doublet: 
\begin{equation}
\mathcal{L}_{V-SM} = 
    - \lambda_{1}^k\, \bar{\psi}_{1L} \tilde{H} u_R^k -
    \lambda_{1d}^k\, \bar{\psi}_{1L} H d_R^k   - \lambda_{2d}^k
    \bar{\psi_2}_L \tilde{H} d^k_R + h.c. \,,  
\end{equation}
where the VLQ fermions are $\psi_1 =({\bf 2},\frac{1}{6})=\left(U_1,D_1\right)^{T}$ and $\psi_2 =
({\bf 2},-\frac{5}{6}) = \left(D_2,Y_2^{-4/3}\right)^T$. The mass Lagrangian is: 
\begin{eqnarray}
\mathcal{L}_{\rm mass} &=& 
   - y_{1}^k \bar{U}_{1L} u_R^k -  y_{1d}^k \bar{D}_{1L} d_R^k - y_{2d}^k \bar{D}_{2L} d_R^k
   - M_1 \, \bar{U}_{1L} U_{1R} -  M_1 \, \bar{D}_{1L} D_{1R}      \nonumber \\ 
&& - M_2 \, \bar{D}_{2L} D_{2R} - M_2 \, \bar{Y}^{-4/3}_{2L} Y^{-4/3}_{2R} + h.c. \,,
\end{eqnarray}
and mass matrices are:
\begin{equation}
M_u = \left( \begin{array}{cc}
\left(\tilde{m}^{up}\right)_{3\times 3} & 0_{3 \times 1} \\
(y_{1}^k)_{1\times 3} & M_1
\end{array} \right)\,, \quad 
M_d = \left( \begin{array}{c c c}
\left(\tilde{m}^{down}\right)_{3\times 3} &  0_{3 \times 1} & 0_{3\times1} \\
(y_{1d}^k)_{1\times 3} & M_1 & 0 \\
(y_{2d}^k)_{1\times 3}  & 0 & M_2
\end{array} \right)\,, \quad 
M_{Y^{-4/3}} = M_2 \,.  
\end{equation}

\subsection{Masses}
\label{app:masses}
In the top-type bi-doublet case we consider,  both VLQ multiplets only couple to the right-handed SM quarks:
\begin{eqnarray}
\mathcal{L}_{V-SM} &=& 
    - \lambda_{1}^k\, \bar{\psi}_{1L} \tilde{H} u_R^k -
    \lambda_{1d}^k\, \bar{\psi}_{1L} H d_R^k - \lambda_2^k \,
    \bar{\psi}_{2L} H u_R^k + h.c. \,,   \label{eq:3:17}
\end{eqnarray}
where $\psi_1 =({\bf 2},\frac{1}{6})=\left(U_1 \ D_1\right)^T$ and $\psi_2 =
({\bf 2},\frac{7}{6}) = \left(X_2^{5/3} \ U_2\right)^T$. 
In this case, no Yukawa coupling between the two VLQ multiplet is allowed, therefore one can use the two free phases to remove one phase in 
$\lambda_1^k$ and one in $\lambda_2^k$, so that only 4 new phases are present in this model. Once again, we will set $\lambda_{1d}^k = 0$.
The mass Lagrangian and mass matrices become:
\begin{eqnarray}
\mathcal{L}_{\rm mass} &=& 
  - y_{1u}^k \bar{U}_{1L} u_R^k -  y_{1d}^k \bar{D}_{1L} d_R^k - y_2^k
  U_{2L} u^k_R                                   \nonumber \\ 
&&  - M_1 \, \bar{U}_{1L} U_{1R} - M_1 \, \bar{D}_{1L} D_{1R} -  M_2
\, \bar{U}_{2L} U_{2R} -  M_2 \, \bar{X}^{5/3}_{2L} X^{5/3}_{2R} + h.c. \,, 
\label{eq:3:18} 
\end{eqnarray}
and
\begin{equation}
M_u = \left( \begin{array}{c c c}
\left(\tilde{m}^{up}\right)_{3\times 3} &  0_{3 \times 1} &
0_{3\times1} \\ 
 (y_{1}^k)_{1\times 3} & M_1 &  0 \\
(y_2^k)_{1 \times 3}  & 0 & M_2
\end{array} \right), \quad
M_d = \left( \begin{array}{c c }
\left(\tilde{m}^{down}\right)_{3\times 3} & 0_{3\times 1}  \\
(0)_{1\times 3} & M_2 \\
\end{array} \right), \quad 
M_{X^{5/3}} = M_2 \,.      \label{eq:3:19}
\end{equation}
The masses of $B^{\prime}$ and $X^{5/3}$ are:
\begin{eqnarray}
m_{b^{\prime}} &=& M_{2} \,,\\
m_{X^{5/3}} &=& M_{2} \,.
\end{eqnarray}
We define the dimensionless $X$ and $Y$ parameters as
\begin{equation}
X=\frac{(y_{1}^{k})^{2}}{M_{1}^{2}-m_{k}^{2}}\,, \quad 
Y=\frac{(y_{2}^{k})^{2}}{M_{2}^{2}-m_{k}^{2}}\,,
\end{equation}
where $k=u,c,t$.
In the bi-doublet model, the top Yukawa coupling and masses of heavy top partners can be written as
\begin{eqnarray}
\tilde{m}_{t}^{2} &=& m_{t}^{2} \left(1 + X + Y \right) \,,\\
m_{t^{\prime}_{1}}^{2} &=& m_{t^{\prime}}^{2} - \frac{\Delta m_{t^{\prime}}^{2}}{2} \,,\\
m_{t^{\prime}_{2}}^{2} &=& m_{t^{\prime}}^{2} + \frac{\Delta m_{t^{\prime}}^{2}}{2} \,,
\end{eqnarray}
with
\begin{eqnarray}
m_{t^{\prime}}^{2} &=& \frac{M_{1}^{2}(1+X) + M_{2}^{2}(1+Y)}{2} \,,\\
\Delta m_{t^{\prime}}^{2}
&=& 2\sqrt{m_{t^{\prime}}^{4} -M_{1}^{2}M_{2}^{2}\frac{\tilde{m}_{t}^{2}}{m_{t}^{2}}} \,.
\end{eqnarray}
For $M_{1}=M_{2}=M$, these can be written as
\begin{eqnarray}
\tilde{m}_{t}^{2} &=& m_{t}^{2} \left( 1 + \frac{(y_{1}^{t})^{2}+(y_{2}^{t})^{2}}{M^{2}-m_{t}^{2}}\right) \,,\\
m_{t^{\prime}_{1}}^{2} &=& M^{2} \,,\\
m_{t^{\prime}_{2}}^{2} &=& M^{2} \left( 1 + \frac{(y_{1}^{t})^{2}+(y_{2}^{t})^{2}}{M^{2}-m_{t}^{2}}\right) \,.
\end{eqnarray}
If $M \gg y_{1}^{t},y_{2}^{t}$:
\begin{eqnarray}
\tilde{m}_{t}^{2} &=& m_{t}^{2} \left( 1 + \frac{(y_{1}^{t})^{2}+(y_{2}^{t})^{2}}{M^{2}} \right) + {\cal O}\left(\frac{1}{M^{4}}\right) \,,\\
\Delta m_{t^{\prime}}^{2} &=& ((y_{1}^{t})^{2}+(y_{2}^{t})^{2}) \left( 1 + \frac{m_{t}^{2}}{M^{2}} \right) + {\cal O}\left(\frac{1}{M^{4}}\right) \,.
\end{eqnarray}
As we showed in the main text, after imposing precision and low energy constraints in the parameter
space of Yukawa couplings, a diagonal band is allowed by both the
constraints even for large Yukawa couplings. When the gauge eigenstate masses of VLQ T-quarks are
degenerate (i.e. $M_1 = M_2$) then $BR(t_1^\prime \to q Z) = 100\%$. This changes when VLT quarks are non-degenerate.

\subsection{Couplings to gauge bosons}
\label{app:couplings}
In the gauge basis, the general expressions for the couplings of $W^{\pm}$ bosons in the two VLQ multiplets models are given by
\begin{eqnarray}
\mathcal{L}_{W^{\pm}} &=&
\frac{g}{\sqrt{2}} \left(
\begin{array}{ccccc}
\bar{u}_{L}^{1}, & \bar{u}_{L}^{2}, & \bar{u}_{L}^{3}, & \bar{U}_{1L}, & \bar{U}_{2L} \\
\end{array}
\right) \cdot \delta_{L} \cdot \gamma^{\mu}
\left(
\begin{array}{c}
d_{L}^{1} \\ d_{L}^{2} \\ d_{L}^{3} \\ D_{1L} \\ D_{2L} \\
\end{array}
\right) W_{\mu}^{+} \nonumber \\
&+& \frac{g}{\sqrt{2}} \left(
\begin{array}{ccccc}
\bar{u}_{R}^{1}, & \bar{u}_{R}^{2}, & \bar{u}_{R}^{3}, & \bar{U}_{1R}, & \bar{U}_{2R} \\
\end{array}
\right) \cdot \delta_{R} \cdot \gamma^{\mu}
\left(
\begin{array}{c}
d_{R}^{1} \\ d_{R}^{2} \\ d_{R}^{3} \\ D_{1R} \\ D_{2R} \\
\end{array}
\right) W_{\mu}^{+} + h.c. \,,
\end{eqnarray}
with
\begin{equation}
\delta_{L} =
\left(
\begin{array}{ccc}
I_{3\times 3} & & \\
& \alpha_{1} & \\
& & \alpha_{2} \\
\end{array}
\right) \,, \quad
\delta_{R} =
\left(
\begin{array}{ccc}
0_{3\times 3} & & \\
& \alpha_{1} & \\
& & \alpha_{2} \\
\end{array}
\right) \,,
\end{equation}
where the values for the $\alpha_i$ coefficients are reported in Table 5 of Ref.~\cite{Cacciapaglia:2015ixa}.
In the mass basis, the left- and right-handed couplings can be written as
\begin{eqnarray}
g_{WL}^{IJ} &=& \frac{g}{\sqrt{2}}V_{CKM}^{L,IJ} = \frac{g}{\sqrt{2}} V_{L}^{u\dagger} \cdot \delta_{L} \cdot V_{L}^{d} \,,\\
g_{WR}^{IJ} &=& \frac{g}{\sqrt{2}}V_{CKM}^{R,IJ} = \frac{g}{\sqrt{2}} V_{R}^{u\dagger} \cdot \delta_{R} \cdot V_{R}^{d} \,,
\end{eqnarray}
where $V_{CKM}^{L}$ and $V_{CKM}^{R}$ are the left- and right-handed CKM matrix, respectively and $V_{L,R}$ are the mixing matrices in the left- and right-handed sectors respectively.

The general expression for the left-handed couplings of the $Z$ in the up quark sector can be written as:
\begin{equation}
\mathcal{L}_{Z} = \frac{g}{c_W}\, \left( \bar{u}_L^1, \bar{u}_L^2, \bar{u}_L^3, \bar{U}_{1L}, \bar{U}_{2L} \right)\cdot \left[ \left( \frac{1}{2} - 
Q_{u} s_W^2 \right) I_{5\times5}
- \Delta T_{3}^{(up)} \right] \gamma^\mu \cdot \left( \begin{array}{c}
u_L^1\\u_L^2\\u_L^3\\U_{1L}\\U_{2L} \end{array} \right) Z_\mu \,,
\end{equation}
with:
\begin{equation}
\Delta T_{3}^{(up)} = \left( \begin{array}{ccc}
0_{3\times3}&&\\ & \Delta T_{3}^{(1,u)} &\\&& \Delta T_{3}^{(2,u)}\\
\end{array} \right) \,,
\end{equation}
where $I_{5\times5}$ is the $5\times5$ unit matrix and $\Delta T_{3}^{(k,u)}=1/2-T_{3}^{(k,u)}$ is the differences between 
the SM top-type quark and $k$-th generation VLQ.
In the  mass eigenstate basis, the left-handed coupling becomes:
\begin{equation}
g_{ZL}^{u,IJ} = \frac{g}{\cw} \left[ \left( \frac{1}{2} - Q_{u} \ssw \right) \delta^{IJ} - \sum_{k=1,2} \Delta T_{3}^{(k,u)} \left( V_L^{u*} \right)^{k+3,I} 
\left( V_L^{u} \right)^{k+3,J} \right] \,.
\end{equation}
Analogously for the right-handed couplings we obtain:
\begin{equation}
g_{ZR}^{u,IJ} = \frac{g}{\cw} \left[ \left(- Q_{u} \ssw \right) \delta^{IJ} + \sum_{k=1,2} T_{3}^{(k,u)} \left( V_R^{u*} \right)^{k+3,I} 
\left( V_R^{u} \right)^{k+3,J} \right] \,.
\end{equation}

In the interaction basis, the Yukawa interactions in top-type quarks can be written as:
\begin{equation}
\mathcal{L}_{H} = \frac{1}{v}\, \left( \bar{u}_L^1, \bar{u}_L^2, \bar{u}_L^3, \bar{U}_{1L}, \bar{U}_{2L}\right)\cdot \left[ M_{u} - M \right]  \cdot \left( \begin{array}{c}
u_R^1\\u_R^2\\u_R^3\\U_{1R}\\U_{2R} \end{array} \right) h + h.c. \,,
\end{equation}
with:
\begin{equation}
M = \left( 
\begin{array}{ccc}
0_{3\times3}&&\\&M_{1}&\\&&M_{2}
\end{array}
\right) \,.
\end{equation}
In the mass eigenstate basis the coupling of top-type quark reads :
\begin{eqnarray}
C_{L}^{u,IJ} &=& \frac{M_{u}^{diag,IJ}}{v} - \sum_{k=1,2}\frac{M_{k}}{v} \left( V_R^{u*} \right)^{k+3,I} \left( V_L^{u} \right)^{k+3,J}\,, \\
C_{R}^{u,IJ} &=& \frac{M_{u}^{diag,IJ}}{v} - \sum_{k=1,2}\frac{M_{k}}{v} \left( V_L^{u*} \right)^{k+3,I} \left( V_R^{u} \right)^{k+3,J}\,.
\label{eq:cptp}
\end{eqnarray}
For bottom-type quark, we obtain:
\begin{eqnarray}
C_{L}^{d,IJ} &=& \frac{M_{d}^{diag,IJ}}{v} - \sum_{k=1,2}\frac{M_{k}}{v} \left( V_R^{d*} \right)^{k+3,I} \left( V_L^{d} \right)^{k+3,J}\,, \\
C_{R}^{d,IJ} &=& \frac{M_{d}^{diag,IJ}}{v} - \sum_{k=1,2}\frac{M_{k}}{v} \left( V_L^{d*} \right)^{k+3,I} \left( V_R^{d} \right)^{k+3,J}\,.
\end{eqnarray}

\subsection{Branching ratios}
\label{app:BRs}
In the top-type bi-doublet case we consider, the VLQ $t^{\prime}_{1}$ and $t^{\prime}_{2}$ ($u_{I=4}$ and $u_{I=5}$) can decay at tree level to $d_{J}W^{+}$ and $X^{5/3}W^{-}$ via a charged current and to $Zu_{J}$ and $hu_{J}$ via a neutral current.
The $u_{I}\to d_{J}W^{+}$ transition matrix element is given by
\begin{equation}
{\cal M} = \bar{d}_{J}(q_{1})\gamma^{\mu} \left( (g_{WL}^{JI})^{*} L + (g_{WR}^{JI})^{*} R \right) u_{I}(q_{2}) \epsilon_{\mu}(\lambda) \,,
\end{equation}
where $L$ and $R$ are the left- and right-handed projection operators.
The partial width of $u_{I}\to d_{J}W^{+}$ decay is expressed as
\begin{eqnarray}
\Gamma(u_{I} \to d_{J}W^{+}) &=& \frac{\lambda^{\frac{1}{2}}(1,m_{d_{J}}^{2}/m_{u_{I}}^{2},m_{W}^{2}/m_{u_{I}}^{2})}{32\pi m_{u_{I}}} \nonumber\\
&& \Biggl\{ \left( |g_{WL}^{JI}|^{2} + |g_{WR}^{JI}|^{2} \right) \left[ m_{u_{I}}^{2} + m_{d_{J}}^{2} - 2m_{W}^{2} + \frac{(m_{u_{I}}^{2}-m_{d_{J}}^{2})^{2}}{m_{W}^{2}} \right] \nonumber\\
&& -12 \left( \re g_{WL}^{JI} \re g_{WR}^{JI} + \im g_{WL}^{JI} \im g_{WR}^{JI} \right) m_{u_{I}}m_{d_{J}} \Biggr\} \,,
\end{eqnarray}
where $I=4,5$ and $J=1,\cdots,4$, $\lambda(a,b,c)=a^2+b^2+c^2-2ab-2ac-2bc$ is the phase space function, and $\re$ and $\im$ indicate the real and imaginary part, respectively.

The $u_{I}\to X^{5/3}W^{-}$ transition matrix element is given by
\begin{equation}
{\cal M} = \bar{X}^{5/3}(q_{1})\gamma^{\mu} \left( g_{WL}^{X^{5/3},I5} L + g_{WR}^{X^{5/3},I5} R \right) u_{I}(q_{2}) \epsilon_{\mu}(\lambda) \,,
\end{equation}
where $g_{WL}^{X^{5/3},I5}$ and $g_{WR}^{X^{5/3},I5}$ are the left- and right-handed couplings of $X^{5/3}$ given in Ref.~\cite{Cacciapaglia:2015ixa}.
The partial width of $u_{I}\to X^{5/3}W^{-}$ decay is expressed as
\begin{eqnarray}
\Gamma(u_{I} \to X^{5/3}W^{-}) &=& \frac{\lambda^{\frac{1}{2}}(1,m_{X^{5/3}}^{2}/m_{u_{I}}^{2},m_{W}^{2}/m_{u_{I}}^{2})}{32\pi m_{u_{I}}} \nonumber\\
&& \Biggl\{ \left( |g_{WL}^{X^{5/3},I5}|^{2} + |g_{WR}^{X^{5/3},I5}|^{2} \right) \left[ m_{u_{I}}^{2} + m_{X^{5/3}}^{2} - 2m_{W}^{2} + \frac{\left(m_{u_{I}}^{2}-m_{X^{5/3}}^{2}\right)^{2}}{m_{W}^{2}} \right] \nonumber\\
&& -12 \left( \re g_{WL}^{X^{5/3},I5} \re g_{WR}^{X^{5/3},I5} + \im g_{WL}^{X^{5/3},I5} \im g_{WR}^{X^{5/3},I5} \right) m_{u_{I}}m_{X^{5/3}} \Biggr\} \,.
\end{eqnarray}

Concerning neural currents, the $u_{I}\to u_{J}Z$ transition matrix element is written as
\begin{equation}
{\cal M} = \bar{u}_{J}(q_{1})\gamma^{\mu} \left( g_{ZL}^{u,IJ} L + g_{ZR}^{u,IJ} R \right) u_{I}(q_{2}) \epsilon_{\mu}(\lambda) \,.
\end{equation}
The partial width of $u_{I}\to u_{J}Z$ is
\begin{eqnarray}
\Gamma(u_{I} \to u_{J}Z) &=& \frac{\lambda^{\frac{1}{2}}(1,m_{u_{J}}^{2}/m_{u_{I}}^{2},m_{Z}^{2}/m_{u_{I}}^{2})}{32\pi m_{u_{I}}} \nonumber\\
&& \Biggl\{ \left( |g_{ZL}^{u,IJ}|^{2} + |g_{ZR}^{u,IJ}|^{2} \right) \left[ m_{u_{I}}^{2} + m_{u_{J}}^{2} - 2m_{Z}^{2} + \frac{(m_{u_{I}}^{2}-m_{u_{J}}^{2})^{2}}{m_{Z}^{2}} \right] \nonumber\\
&& -12 \left( \re g_{ZL}^{u,IJ} \re g_{ZR}^{u,IJ} + \im g_{ZL}^{u,IJ} \im g_{ZR}^{u,IJ} \right) m_{u_{I}}m_{u_{J}} \Biggr\} \,,
\end{eqnarray}
where $I=4,5$ and $J=1,2,\cdots<I$.

The matrix element of $u_{I}\to u_{J}h$ is written as
\begin{equation}
{\cal M} = \bar{u}_{J}(q_{1}) \left( C_{L}^{u,IJ} L + C_{R}^{u,IJ} R \right) u_{I}(q_{2}) \,.
\end{equation}
The partial width of $u_{I}\to u_{J}h$
\begin{eqnarray}
\Gamma(u_{I} \to u_{J}h) &=& \frac{\lambda^{\frac{1}{2}}(1,m_{u_{J}}^{2}/m_{u_{I}}^{2},m_{h}^{2}/m_{u_{I}}^{2})}{32\pi m_{u_{I}}} \nonumber\\
&& \Biggl[ (m_{u_{I}}^{2}+m_{u_{J}}^{2}-m_{h}^{2}) \left( |C_{L}^{u,IJ}|^{2} + |C_{R}^{u,IJ}|^{2} \right) + 4m_{u_{I}}m_{u_{J}} \left( \re C_{L}^{u,IJ} \re C_{R}^{u,IJ}  \right. \nonumber \\ 
&& \left. + \im C_{L}^{u,IJ} \im C_{R}^{u,IJ} \right) \Biggr] \,,
\end{eqnarray}
where $I=4,5$ and $J=1,2,\cdots<I$.

The total decay width of $u_{I}$ is given by
\begin{eqnarray}
\Gamma_{total}(u_{I}) &=& \Gamma(u_{I} \to X^{5/3}W^{-}) + \sum_{J=1}^{4} \Gamma(u_{I} \to d_{J}W^{+}) \nonumber\\
&& + \sum_{I>J} \left( \Gamma(u_{I} \to u_{J}Z) + \Gamma(u_{I} \to u_{J}h) \right) \,.
\end{eqnarray}


\bibliographystyle{JHEP}
\bibliography{VLQnew}


\end{document}